\documentclass{ptephy_v1}

\preprintnumber{XXXX-XXXX} 

\usepackage{txfonts}
\usepackage{amsmath,amsfonts,amssymb,amsthm}
\usepackage{mathrsfs}
\usepackage{graphicx}
\usepackage{braket}
\usepackage{comment}
\usepackage{enumerate}

\usepackage{epic,eepic}

\numberwithin{equation}{section}

\begin{document}

\title{Exhaustive derivation of static self-consistent multi-soliton solutions in \\ the matrix Bogoliubov-de~Gennes systems}


\author{Daisuke A. Takahashi}
\affil{RIKEN Center for Emergent Matter Science (CEMS), Wako, Saitama 351-0198, Japan \\ Research and Education Center for Natural Sciences, Keio University, Hiyoshi 4-1-1, Yokohama, Kanagawa 223-8521, Japan \email{daisuke.takahashi.ss@riken.jp}}





\begin{abstract}%
The matrix-generalized Bogoliubov-de Gennes systems have recently been considered by the present author [Phys. Rev. B \textbf{93}, 024512 (2016)], and time-dependent and self-consistent multi-soliton solutions have been constructed based on the ansatz method. In this paper, restricting the problem to the static case, we exhaustively determine the self-consistent solutions using the inverse scattering theory. Solving the gap equation, we rigorously prove that the self-consistent potential must be reflectionless. As a supplementary topic, we elucidate the relation between the stationary self-consistent potentials and the soliton solutions in the matrix nonlinear Schr\"odinger equation. Asymptotic formulae of multi-soliton solutions for sufficiently isolated solitons are also presented.
\end{abstract}

\subjectindex{A11, B34, I20, I22, I68}

\maketitle

\section{Introduction}
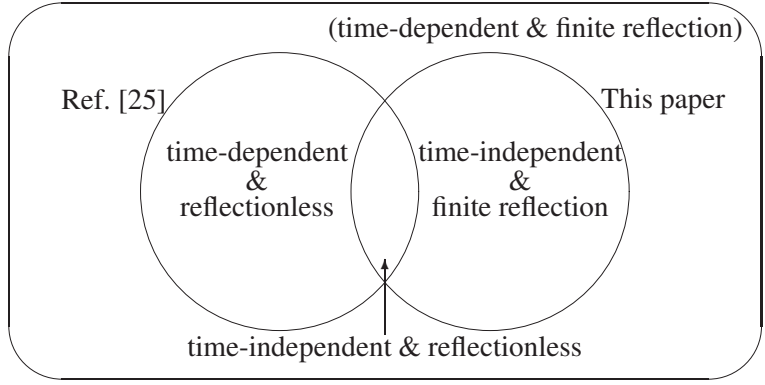
\begin{figure}[tb]
		\begin{center}
		{\fontsize{11pt}{24pt}\selectfont
		\setlength\unitlength{1cm}
		\begin{picture}(9,5)(1,0)
		\put(3.6,2.5){\circle{3.7}}
		\put(6.4,2.5){\circle{3.7}}
		\put(5,2.5){\oval(10,5)}
		\put(5,0.6){\vector(0,1){1}}
		\put(5,0.4){\makebox(0,0)[c]{time-independent \& reflectionless}}
		\put(1.4,3.7){\makebox(0,0)[c]{Ref.~\cite{arxiv1509.04242}}}
		\put(3.3,3){\makebox(0,0)[c]{time-dependent}}
		\put(3.3,2.65){\makebox(0,0)[c]{\&}}
		\put(3.3,2.3){\makebox(0,0)[c]{reflectionless}}
		\put(8.7,3.7){\makebox(0,0)[c]{This paper}}
		\put(6.8,3){\makebox(0,0)[c]{time-independent}}
		\put(6.8,2.65){\makebox(0,0)[c]{\&}}
		\put(6.8,2.3){\makebox(0,0)[c]{finite reflection}}
		\put(7,4.65){\makebox(0,0)[c]{(time-dependent \& finite reflection)}}
		\end{picture}
		}
		\caption{\label{fig:intro} The relation between Ref.~\cite{arxiv1509.04242} and this paper. The two works have a common subset, i.e., the time-independent and reflectionless solution. It must be emphasized that, though we fully consider potentials possessing finite reflection coefficients, we eventually conclude that \textit{the self-consistent potential must be reflectionless} (Sec.~\ref{sec:gap}). 
		As for the one-component problem, the time-independent and reflectionless case was solved in Ref.~\cite{PhysRevLett.110.131601}, the time-independent and finite-reflection case was treated in Ref.~\cite{takahashinittaJLTP} (several earlier discussions of the reflectionless nature can also be found in \cite{Dashen:1975xh,Shei:1976mn,Feinberg:2003qz}), and time-dependent reflectionless solutions were constructed in Refs.~\cite{PhysRevLett.111.121602,PhysRevA.88.062115,PhysRevD.89.025008}.}
		\end{center}
	\end{figure}

	The Bogoliubov-de Gennes (BdG) equation \cite{Bogoliubov1958,DeGennes:1999} and the associated self-consistent condition (the gap equation) are fundamental equations in the mean-field treatment 
	of superconductors and fermionic superfluids, and the natures of Bogoliubov quasiparticles and order parameters (gap functions) are determined by them. 
	It also describes many other condensed-matter systems including conducting polymers such as polyacetylene and chromium alloys \cite{Su:1979ua,Takayama:1980zz,RevModPhys.60.781,MachidaFujita,Fawcett}. 
	Equivalent problems are also known to appear in high-energy physics in the Hartree-Fock theory of the Nambu-Jona Lasinio and Gross-Neveu models \cite{Nambu:1961tp,Gross:1974jv,Dashen:1975xh}. 
	While it is generally difficult to obtain an analytical solution satisfying both the BdG and the gap equations, soliton-theoretical techniques can be applied in one spatial dimension and many exact multi-soliton solutions have been obtained so far \cite{Dashen:1975xh,Shei:1976mn,Takayama:1980zz,OkunoOnodera,OkunoOnodera2,Feinberg:2003qz,Feinberg:2002nq,PhysRevD.83.085001,Correa:2009xa,PhysRevB.84.024503,Takahashi:2012aw,PhysRevLett.110.131601,PhysRevLett.111.121602,PhysRevA.88.062115,PhysRevD.89.025008,Arancibia:2014lwa}. \\  
	\indent In Ref.~\cite{arxiv1509.04242}, the present author generalized the BdG model to the multicomponent case. Here we call it the matrix BdG model.  The matrix BdG model includes examples of unconventional superconductors such as triplet $p$-wave superfluids/superconductors in condensed matter and $SU(n)$-symmetric fermions in ultracold atoms (see, e.g., Refs.~\cite{VollhardtWolfle,JPSJ.81.011009,0953-8984-27-11-113203,arxiv1508.00787,PhysRevLett.98.030401,JPSJ.78.012001,0034-4885-77-12-124401} and references therein). 
	The author constructed the time-dependent soliton solutions based on the ansatz originating from the Gelfand-Levitan-Marchenko (GLM) equation in the inverse scattering theory (IST). In particular the  $ 2\times 2 $ case, which describes a mixture of triplet $p$-wave and singlet $s$-wave superconductors/superfluids, is investigated in detail and various examples of multi-soliton dynamics are demonstrated. 
	 The related problem in high-energy physics has been considered more recently \cite{arxiv1512.03894}. \\ 
	\indent In this paper, we attack the problem from another direction. Restricting the problem to the static one, we exhaustively solve it without relying on an ansatz. Using the IST, we take into account all kinds of potentials having nonzero reflection coefficients, and we prove rigorously that the self-consistent solution must be reflectionless. The relation between this paper and Ref.~\cite{arxiv1509.04242} is depicted in Fig.~\ref{fig:intro}.\\ 
	\indent In fact, the time-dependent multi-component solutions reported in Ref.~\cite{arxiv1509.04242} were obtained by ``extrapolating'' the static multi-component solutions given in the present paper and the time-dependent one-component solutions given in Refs.~\cite{PhysRevLett.111.121602,PhysRevA.88.062115,PhysRevD.89.025008}. 
	We believe that the mathematical expressions and techniques provided in this work will offer an insight into the most general time-dependent and finite-reflection solutions shown in Fig.~\ref{fig:intro}. \\
	\indent Here we explain the organization of this paper with a compact summary of each section. Sections~\ref{sec:ist} and \ref{sec:gap} are devoted to the main topic of this paper. In Sec.~\ref{sec:ist}, we formulate the IST of the matrix Zakharov-Shabat (ZS) operator with finite-density boundary condition. While the same problem is also considered in the context of solving the matrix nonlinear Schr\"odinger (NLS) equation \cite{jmp48110.10631.2423222}, in order to solve the gap equation, we need several new additional formulae and techniques, such as the treatment of degenerate eigenvalues and the expression of corresponding bound states (Subsec.~\ref{subsec:boundstates}), the orthogonal and completeness relations (Subsecs.~\ref{eq:kerneltojost}-\ref{subsec:ortho2}), and an inner-product formula for a higher-order product (Appendix~\ref{app:fourortho}). In Sec.~\ref{sec:gap}, we define the gap equations and solve them. We prove that the self-consistent solution must have a reflectionless potential, and the filling rates of bound states become the same as Ref.~\cite{arxiv1509.04242}. Sections \ref{sec:asymptotics} and \ref{sec:matrixnls} provide additional but important topics. 
	In Sec.~\ref{sec:asymptotics}, we present an asymptotic formula for the reflectionless solution, i.e., the $n$-soliton solution, when solitons are well isolated. 
	In Sec.~\ref{sec:matrixnls}, we clarify that the stationary-class potentials are snapshots of each time of the soliton solutions of the matrix NLS equation, as described in Sec. II D of Ref.~\cite{arxiv1509.04242}.
	Section~\ref{sec:summary} gives a summary. Appendices~\ref{app:Kintegral}-\ref{app:fourortho} provide a few mathematical proofs and formulae to support the mathematical rigor.
	
\section{Inverse scattering theory}\label{sec:ist}
\subsection{Matrix Zakharov-Shabat eigenvalue problem with unitary background}
	\indent We use the following notations for the matrix-generalized Pauli matrices:
	\begin{align}
		\sigma_1=\begin{pmatrix} & I_d \\ I_d & \end{pmatrix},\quad \sigma_2=\begin{pmatrix} & -\mathrm{i} I_d \\ \mathrm{i}I_d & \end{pmatrix},\quad \sigma_3=\begin{pmatrix}  I_d & \\ & -I_d \end{pmatrix},
	\end{align}
	where $ I_d $ is a  $ d\times d $ unit matrix. \\
	\indent Let $ \Delta(x) $ be a $ d\times d $ matrix. We consider the following matrix-generalized ZS eigenvalue problem:
	\begin{align}
		\epsilon w = \mathcal{L}w,\quad \mathcal{L}=\begin{pmatrix} -\mathrm{i}\partial_x I_d & \Delta(x) \\ \Delta(x)^\dagger & \mathrm{i}\partial_x I_d \end{pmatrix}. \label{eq:mZS}
	\end{align}
	We sometimes write $ w=\left( \begin{smallmatrix}u \\ v \end{smallmatrix} \right) $, where $ u $ and $ v $ are $ d $-component vectors. Here, we assume that the asymptotic form of  $ \Delta(x) $ is given by
	\begin{align}
		\Delta(x\rightarrow\pm\infty) = m \Delta_\pm,\quad m>0,\quad  \Delta_\pm: \text{ unitary}. \label{eq:finitebg}
	\end{align}
	This means that the singular values of  $ \Delta(x) $ at infinity are all  $ m $. If we consider the application to physical systems for the $ d=2 $ case,  the unitary background means that the phase of the background order parameter (e.g., spin-1 Bose condensates \cite{JPSJ.75.064002} for a bosonic example or superfluid helium 3 \cite{arxiv1509.04242} for a fermionic example) is non-magnetic. 
\subsection{Symmetric and antisymmetric cases and complex conjugate solution}
	\indent In addition to the general non-symmetric $ \Delta(x) $, we also consider the special cases where $ \Delta(x) $ has the symmetry
	\begin{align}
		&\Delta(x)=D_1\Delta(x)^T D_2^\dagger,\quad  D_1,D_2 \text{ are unitary.}  \label{eq:d1d2sym} \\
		\leftrightarrow\quad& \mathcal{L}=\tau \mathcal{L}^*\tau^\dagger,\quad \tau:=\begin{pmatrix} & D_1 \\ D_2 & \end{pmatrix}.
	\end{align}
	When we write down expressions valid only under these symmetries, we always write ``$ (\text{if }\mathcal{L}=\tau \mathcal{L}^*\tau^\dagger) $''.
	Note that $ \Delta_\pm=D_1 \Delta_\pm D_2^\dagger $ also holds, since the relation must be satisfied also for $ x=\pm\infty $. 
	Since the above relation can be rewritten as $ \Delta(x)=D_2^T\Delta(x)^TD_1^* $, the following relation holds:
	\begin{align}
		D_2=\pm D_1^T,\quad \therefore \tau=\pm\tau^T \quad(\leftrightarrow \tau^\dagger=\pm \tau^*). \label{eq:taucondition}
	\end{align}
	Moreover, using the global unitary transformation $ (u,v,\Delta)\rightarrow (Uu,Vv,U\Delta V^{-1}) $, we can always set $ D_1 \propto I_d $, which implies that the symmetry of Eq.~(\ref{eq:d1d2sym}) can be always reduced to $ \Delta(x)=\pm\Delta(x)^T $. However, for convenience, we use the redundant notations $ D_1 $ and $ D_2 $. \\
	\indent We can soon check the following complex conjugate solution:
	\begin{align}
		\mathcal{L}w=\epsilon w \quad \leftrightarrow \quad \mathcal{L}\tau w^*=\epsilon^*\tau w^*, \quad (\text{If }\mathcal{L}=\tau \mathcal{L}^*\tau^\dagger.)
	\end{align}
	In particular, the symmetric or antisymmetric cases $ \Delta(x)=\pm\Delta(x)^T $ are realized by 
	\begin{align}
		\Delta=\Delta^T:&\quad D_1=D_2=I_d,\quad \tau=\sigma_1, \\
		\Delta=-\Delta^T:&\quad D_1=-D_2=-\mathrm{i}I_d,\quad \tau=\sigma_2.
	\end{align}
	If we consider the physical problem of fermionic superfluids or superconductors, the antisymmetric case describes $ s $-wave superfluids, and the symmetric case describes $ p $-wave superfluids \cite{arxiv1509.04242}. 
	Here, we note that the differential operators in the off-diagonal terms of the $p$-wave BdG equation, having a form like $ \{-\mathrm{i}\partial, \Delta \} $, are now replaced by $ k_F $ by the Andreev approximation, i.e., the dispersion linearization around the Fermi point. In this treatment, $p$-wave order parameters are simply described by a symmetric matrix without differential operators. A new dimensionless unit with $k_F=1$ is taken after this approximation. See also Sec. IV of Ref.~\cite{arxiv1509.04242} for details of the formulation.
\subsection{Uniformization variable}\label{subsec:univar}
	\indent The uniformization variable (\u{Z}ukowsky transform \cite{FaddeevTakhtajan}) is defined in the same way as before \cite{PhysRevLett.110.131601,arxiv1509.04242}:
	\begin{align}
		\epsilon(s)=\frac{m}{2}(s+s^{-1}),\ k(s)=\frac{m}{2}(s-s^{-1}).
	\end{align}
	By this parametrization, the dispersion relation for the uniform system $ \epsilon^2=k^2+m^2 $ holds automatically. As explained in Ref.~\cite{PhysRevLett.110.131601},  $ s\in \mathbb{R} $ corresponds to the scattering states, and $ s \in \mathbb{H} \ \& \ |s|=1 $ corresponds to the bound states. 
	If we consider time-dependent solutions,  $ s\in \mathbb{H} \ \& \ |s| \lessgtr 1 $ corresponds to moving solitons with velocity $ V=\frac{1-|s|^2}{1+|s|^2} \gtrless 0 $ \cite{PhysRevLett.111.121602,PhysRevA.88.062115,PhysRevD.89.025008,arxiv1509.04242}.
\subsection{Constants and orthogonal relations}\label{subsec:constantWJ}
	Let $ f_1,\dots,f_{2d} $ be solutions of Eq.~(\ref{eq:mZS}) for the same $ \epsilon $. Then, the Wronskian defined by
	\begin{align}
		W=\det(f_1,\dots, f_{2d})
	\end{align}
	is a constant, i.e.,  $ W $ does not depend on $ x $. In particular,  $ W\ne 0 $ means that $ f_1,\dots,f_{2d} $ are linearly independent; thus, they become one basis for the solution space of Eq.~(\ref{eq:mZS}) for given $ \epsilon $. \\
	\indent Let $ w_i= w_i(x,s_i) $ be a solution of Eq. (\ref{eq:mZS}) with  $ \epsilon_i=\epsilon(s_i) $, and assume that it is analytic with respect to $ s_i $ in some subset of $ \mathbb{C} $. Then, $ w_i(x,s_i)^\dagger $ is an analytic function of $ s_i^* $. Henceforth we show the $x$-derivative of $ f $ by a subscript $ f_x $ and the $s$-derivative by dot $ \dot{f} $. From Eq. (\ref{eq:mZS}), we soon find
	\begin{align}
		(w_i^\dagger\sigma_3 w_j)_x=-\mathrm{i}(\epsilon_i^*-\epsilon_j)w_i^\dagger w_j, \label{eq:constwjnew01}
	\end{align}   
	which represents the orthogonality of eigenstates between different eigenvalues. In particular, when $ \epsilon_i^*=\epsilon_j $, 
	\begin{align}
		J=w_i^\dagger\sigma_3 w_j = \text{const}. \quad (\text{if } \epsilon_i^*=\epsilon_j).
	\end{align}
	The constancy of $ J $ is used to define the transmission and reflection coefficients (Subsec.~\ref{subsec:jostscat}). Differentiating Eq. (\ref{eq:constwjnew01}) by  $ s_j $ and setting  $ \epsilon_i^*=\epsilon_j $, 
	\begin{align}
		(w_i^\dagger\sigma_3\dot{w}_j)_x=\mathrm{i}\dot{\epsilon}_jw_i^\dagger w_j \quad (\text{if } \epsilon_i^*=\epsilon_j).  \label{eq:boundprod}
	\end{align}
	From Eq.~(\ref{eq:constwjnew01}), we also obtain
	\begin{align}
		(w_k^\dagger \sigma_3 w_i w_j^\dagger \sigma_3 w_l)_x=-\mathrm{i}(\epsilon_k^*-\epsilon_i)w_k^\dagger[w_iw_j^\dagger,\sigma_3]w_l \quad (\text{if } \epsilon_l=\epsilon_k^* \text{ and } \epsilon_j=\epsilon_i^*). \label{eq:constwjnew02}
	\end{align}
	Equation (\ref{eq:boundprod}) is used to calculate a normalization constant of bound states in Subsec.~\ref{subsec:boundstates} and Eq.~(\ref{eq:constwjnew02}) is used to solve the gap equation in Subsec.~\ref{subsec:solvegap}.
\subsection{Solutions for uniform systems}
	Let us consider the uniform  $ \Delta(x)=m\Delta_{\pm} $ in Eq.~(\ref{eq:mZS}). For given $ \epsilon=\epsilon(s) \ (s\ne\pm1) $, the linearly independent $ 2d $ eigenstates are given by
	\begin{align}
		\Phi_{\pm}(x,s):=\begin{pmatrix} s \Delta_\pm \\ I_d \end{pmatrix}\mathrm{e}^{\mathrm{i}k(s)x}, \quad s\Phi_{\pm}(x,s^{-1}) =\begin{pmatrix} \Delta_\pm \\ sI_d \end{pmatrix}\mathrm{e}^{-\mathrm{i}k(s)x}.
	\end{align}
	Each column of $ \Phi $ satisfies Eq.~(\ref{eq:mZS}); thus, the above provides $ 2d $ solutions. We also define the $ 2d \times 2d $ matrix solution by
	\begin{align}
		&\Psi_\pm(x,s):=\Bigl[ \Phi_{\pm}(x,s)\Delta_\pm^\dagger,\ s\Phi_{\pm}(x,s^{-1})  \Bigr]=(sI_{2d}+M_\pm)\mathrm{e}^{\mathrm{i}k(s)x\sigma_3}, \label{eq:Psiintroduce}  \\
		&M_\pm:=\begin{pmatrix}  & \Delta_\pm \\ \Delta_\pm^\dagger &  \end{pmatrix}.
	\end{align}
	 $ M_\pm $ is both unitary and hermitian $ M_\pm=M_\pm^\dagger=M_\pm^{-1} $ and anticommutes with $ \sigma_3 $:  $ \{M_\pm,\sigma_3\}=0 $. 
	By definition, the following relation holds:
	\begin{align}
		s\Psi_\pm(x,s^{-1})=\Psi_\pm(x,s)M_\pm. \label{eq:uniformpsi03}
	\end{align}
	The constants $ W $ and $ J $ introduced in Subsec~\ref{subsec:constantWJ} are calculated as 
	\begin{align}
		W&=\det \Psi_\pm(x,s) =(s^2-1)^d, \\
		J&=\Psi_\pm(x,s^*)^\dagger \sigma_3 \Psi_\pm(x,s)=(s^2-1)\sigma_3. \label{eq:constsolJ}
	\end{align}
	The complex conjugation relation for symmetric/antisymmetric cases ($ \mathcal{L}=\tau \mathcal{L}^*\tau^\dagger $) is given by
	\begin{align}
		\tau \Psi_\pm(x,s)^*=\Psi_\pm(x,s^*)\tau, \quad (\text{If }\mathcal{L}=\tau \mathcal{L}^*\tau^\dagger). \label{eq:uniformpsicc}
	\end{align}
\subsection{Jost functions and scattering matrix}\label{subsec:jostscat}
	\indent Henceforth, we consider non-uniform $ \Delta(x) $ with the boundary condition Eq.~(\ref{eq:finitebg}). We define the right ($+$) and left ($-$) Jost functions $ F_\pm(x,s) $  by the solution of Eq. (\ref{eq:mZS}) with the following asymptotic behavior:
	\begin{align}
		F_\pm(x,s) \rightarrow \Phi_\pm(x,s) \quad (x\rightarrow\pm\infty). 
	\end{align}
	We also define the $ 2d\times 2d $ Jost function by $ Y_\pm(x,s):=\Bigl[ F_{\pm}(x,s)\Delta_\pm^\dagger,\ sF_{\pm}(x,s^{-1})  \Bigr] $, which has the asymptotic form 
	\begin{align}
		Y_\pm(x,s) \rightarrow \Psi_\pm(x,s) \quad (x\rightarrow\pm\infty). 
	\end{align}
	Since a Jost function satisfying a given asymptotic form at  $ x=+\infty $ or $ -\infty $ is uniquely fixed (due to the uniqueness of the solution of the linear differential equation for a given initial condition), the Jost functions with the same asymptotic form must be identified as the same one. Therefore, from Eqs. (\ref{eq:uniformpsi03}) and (\ref{eq:uniformpsicc}), we obtain
	\begin{align}
		sY_\pm(x,s^{-1})&=Y_\pm(x,s)M_\pm, \label{eq:jostystosi} \\
		\tau Y_\pm(x,s)^*&=Y_\pm(x,s^*)\tau \quad (\text{if }\mathcal{L}=\tau \mathcal{L}^*\tau^\dagger). \label{eq:jostytau}
	\end{align}
	We introduce a $ 2d\times 2d $ scattering matrix $ S(s) $ by
	\begin{align}
		Y_+(x,s)=Y_-(x,s)S(s). \label{eq:defSmat}
	\end{align}
	Then, by definition, the asymptotic form of $ Y_+(x,s) $ is given by: 
	\begin{align}
		Y_+(x,s) \rightarrow \begin{cases} \Psi_-(x,s)S(s) & (x\rightarrow-\infty) \\ \Psi_+(x,s) & (x\rightarrow+\infty). \end{cases} \label{eq:asymptoticY}
	\end{align}
	Evaluating the constants $ W $ and $ J $ of $ Y_+(x,s) $ at $ x=\pm\infty $, and equating them, we obtain
	\begin{align}
		\det S(s)&=1, \\
		S(s^*)^\dagger\sigma_3S(s)&=\sigma_3. \label{eq:Smatrixinverse}
	\end{align}
	Using Eq.~(\ref{eq:jostystosi}) for the plus sign at $ x=-\infty $ and Eq.~(\ref{eq:uniformpsi03}) for the minus sign, we find
	\begin{align}
		S(s)M_+=M_-S(s^{-1}). \label{eq:scatMpMm}
	\end{align}
	Therefore,  $ S(s) $ has the following form:
	\begin{align}
		S(s)=\begin{pmatrix} A(s) & \Delta_-B(s^{-1})\Delta_+ \\ B(s) &\Delta_-^\dagger A(s^{-1})\Delta_+ \end{pmatrix}, \label{eq:SsAsBs}
	\end{align}
	where $ A(s) $ and $ B(s) $ are $ d\times d $ matrices. \\
	\indent Equation (\ref{eq:Smatrixinverse}) means that the inverse of $ S(s) $ is given by $ S(s)^{-1}=\sigma_3S(s^*)^\dagger\sigma_3 $. Using $ A(s) $ and $ B(s) $, the relation $ S(s)^{-1} S(s)=S(s)S(s)^{-1}=I_{2d} $ is explicitly written as
	\begin{align}
		A(s^*)^\dagger A(s)-B(s^*)^\dagger B(s)&=I_d, \label{eq:scatinv15} \\
		A(s^*)^\dagger \Delta_- B(s^{-1})-B(s^*)^\dagger \Delta_-^\dagger A(s^{-1})&=0, \label{eq:scatinv2} \\
		A(s)A(s^*)^\dagger-\Delta_- B(s^{-1})B(s^{-1*})^\dagger\Delta_-^\dagger&=I_d,  \label{eq:scatinv32} \\
		A(s)B(s^*)^\dagger-\Delta_-B(s^{-1})A(s^{-1*})^\dagger\Delta_-&=0. \label{eq:scatinv42}
	\end{align}
	If we define $ T(s)=A(s)^{-1} $ and $ R(s)=B(s)A(s)^{-1} $, Eqs.~(\ref{eq:scatinv15}) and (\ref{eq:scatinv2}) are rewritten as
	\begin{align}
		T(s^*)^\dagger T(s)+R(s^*)^\dagger R(s)&=I_d, \label{eq:scatinv1501} \\
		\Delta_-R(s^{-1})-R(s^*)^\dagger\Delta_-^\dagger&=0. \label{eq:scatinv202}
	\end{align}
	The former relation suggests that $ T(s) $ and $ R(s) $ are regarded as  transmission and reflection coefficients. 
	In fact, the asymptotic relation (\ref{eq:asymptoticY}) is rewritten as
	\begin{align}
		F_+(x,s)\Delta_+^\dagger &\rightarrow \begin{cases} \Phi_-(x,s)\Delta_-^\dagger A(s)+s\Phi_-(x,s^{-1})B(s) & (x\rightarrow-\infty) \\ \Phi_+(x,s)\Delta_+^\dagger & (x\rightarrow+\infty).  \end{cases} \label{eq:asymptoticF1} 
	\end{align}
	If we multiply Eq.~(\ref{eq:asymptoticF1}) by $ A(s)^{-1} $ from right, and rewrite the expression with $ T(s) $ and $ R(s) $ instead of $ A(s) $ and $ B(s) $, it gives the solution of the transmission-and-reflection problem. \\
	\indent Next, let us consider the complex conjugation relations of the scattering matrix $ S(s) $ when the symmetric or antisymmetric relation $ \mathcal{L}=\tau \mathcal{L}^*\tau^\dagger $ holds. We find 
	\begin{align}
		\tau S(s)^*=S(s^*)\tau \quad (\text{if }\mathcal{L}=\tau \mathcal{L}^*\tau^\dagger), \label{eq:Smatrixtau}
	\end{align}
	by using Eq.~(\ref{eq:jostytau}) with a plus sign at $ x=-\infty $. Thus, Eq.~(\ref{eq:Smatrixinverse}) reduces to 
	\begin{align}
		S(s)^T\tau^\dagger\sigma_3S(s)=\tau^\dagger\sigma_3 \quad (\text{if }\mathcal{L}=\tau \mathcal{L}^*\tau^\dagger). \label{eq:Smatrixtau2}
	\end{align}
	Here we have used $ \tau^\dagger=\pm \tau^* $ [Eq.~(\ref{eq:taucondition})]. This means that $ S(s) \in O(\tau^\dagger \sigma_3) $, where $ O(\sigma) $ denotes a generalized orthogonal group whose element $ X $ satisfies $ X^T\sigma X=\sigma  $. \\ 
	\indent In particular, when $ \tau=\sigma_1 $ and $  \Delta(x)=\Delta(x)^T $, Eqs.~(\ref{eq:scatinv2}) and (\ref{eq:Smatrixtau}) reduce to
	\begin{align}
		A(s)&=\Delta_-A(s^{-1*})^*\Delta_+^\dagger,\\
		B(s)&=\Delta_-^\dagger B(s^{-1*})^*\Delta_+^\dagger, \\
		A(s)^TB(s)&=B(s)^TA(s). \quad (\leftrightarrow R(s)^T=R(s).) \label{eq:absym}
	\end{align}
	On the other hand, when $ \tau=\sigma_2 $ and $ \Delta(x)=-\Delta(x)^T $, Eqs.~(\ref{eq:scatinv2}) and (\ref{eq:Smatrixtau}) reduce to
	\begin{align}
		A(s)&=\Delta_-A(s^{-1*})^*\Delta_+^\dagger,\\
		B(s)&=-\Delta_-^\dagger B(s^{-1*})^*\Delta_+^\dagger, \\
		A(s)^TB(s)&=-B(s)^TA(s). \quad (\leftrightarrow R(s)^T=-R(s).) \label{eq:abasym}
	\end{align}
	Thus, the reflection coefficient $ R $ is symmetric/antisymmetric when $ \Delta $ is symmetric/antisymmetric.

\subsection{Bound states}\label{subsec:boundstates}
In this subsection, we summarize the concept of bound states and zeros of the scattering matrix.
\subsubsection{$C_j$: the residue of $A(s)^{-1}$}
	\begin{figure}[tb]
		\begin{center}
		\includegraphics[scale=.6]{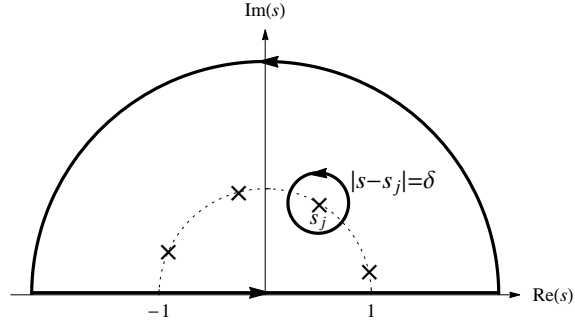}
		\caption{\label{fig:contours} A contour $ |s-s_j|=\delta $ used to define  $ C_j $ [Eq.~(\ref{eq:DefCj})]. Crosses marks represent the zeros of $ \det A(s) $. The large semicircular contour is used to derive the GLM equation in Subsec.~\ref{subsec:glm}.} 
		\end{center}
	\end{figure}
	Normalizable bound states, corresponding to the discrete spectrum, can emerge at the points such that $ \det A(s)=0,\ s\in \mathbb{H} $, because the divergent component in the asymptotic form (\ref{eq:asymptoticF1}) can be eliminated. Since the matrix ZS operator (\ref{eq:mZS}) is self-adjoint, the possible eigenvalue of bound states is restricted to real numbers; thus the zeros of $ \det A(s) $ can appear only for $ |s|=1 $ with $ s \in \mathbb{H} $. \\
	\indent Let $ s_j $ be a zero of $ \det A(s) $ on $ |s|=1 $ and $ s \in \mathbb{H} $. Since $ \det A(s_j)=0 $, the rank of $ A(s_j) $ is less than  $ d $. Let us write $ \operatorname{rank}A(s_j)=d-r $ with $ 1\le r \le d $, and hence $ \dim \operatorname{Ker}A(s_j)=r $. There are $ r $ linearly independent vectors $ c_1,\dots, c_r $ such that $ A(s_j)c_r=0 $, and hence we have $ r $ bound states. As we will prove later,  $ s_j $ is the zero of $ \det A(s) $ of order $ r $, which means that the zeros of matrix eigenvalues of $ A(s) $ are all simple. We want to prepare a matrix $ C_j $ of $ \operatorname{rank}r $ including all $ r $ column vectors of the null space of $ A(s_j) $, which satisfies $ A(s_j)C_j=0 $. If such  $ C_j $ is found, the right Jost function (\ref{eq:asymptoticF1}) with $ s=s_j $ multiplied by $ C_j $, i.e.,  $ F_+(x,s_j)\Delta_+^\dagger C_j $, includes all $r$ linearly independent bound states of eigenvalue $ \epsilon(s_j) $ in its $ d $ columns. 
	 Such $ C_j $ can be constructed as follows:
	\begin{align}
		C_j=\frac{1}{2\pi\mathrm{i}}\int_{|s-s_j|=\delta}\mathrm{d}s\frac{\mathrm{d}[A(s)^{-1}]}{\mathrm{d}s}, \label{eq:DefCj}
	\end{align}
	where $ \delta>0 $ is taken to be sufficiently small to exclude other discrete eigenvalues (See Fig.~\ref{fig:contours}). Below, we explain that $ C_j $ defined by Eq.~(\ref{eq:DefCj}) has a desired property for the above-mentioned purpose.\footnote{
	The cofactor matrix $ C $, which satisfies $ AC=CA=(\det A)I_d $, cannot be used for the current purpose unless $ s=s_j $ is a simple zero, because the rank of the cofactor matrix is generally given by $ \operatorname{rank}C=d,\,1,\ $ and $0$ when $ \operatorname{rank}A=d,\ d-1,\ $ and otherwise. (What we want here is a matrix such that $ \operatorname{rank}A+\operatorname{rank}C=d $ and $ AC=CA=0 $.)
	}
	\\
	\indent By the singular-value decomposition (SVD) theorem, there exist  $ s $-independent unitary matrices $ \mathcal{U} $ and $ \mathcal{V} $ such that $ A(s) $ in the vicinity of $ s=s_j $ can be expressed as
	\begin{align}
		A(s)=\mathcal{U}\left[ \begin{pmatrix} O_r & \\ & \Lambda \end{pmatrix} + (s-s_j)\begin{pmatrix} \tilde{\Lambda} & * \\ * & * \end{pmatrix}  \right]\mathcal{V}^\dagger+O((s-s_j)^2), \label{eq:Assvd}
	\end{align}
	where $ O_r $ is a zero matrix of size $ r $, and $ \Lambda $ is a $ (d-r)\times (d-r) $ diagonal matrix whose diagonal elements are real and positive by assumption.  $ \tilde{\Lambda} $ is an $ r\times r $ diagonal matrix whose diagonal elements are real and non-negative, which is in fact proved to be positive later (footnote  \ref{foot1}). Therefore both $ \Lambda $ and $ \tilde{\Lambda} $ are invertible. The asterisk represents some nonzero matrix not important in the current problem. The existence of the expression (\ref{eq:Assvd}) is shown as follows. First, make an SVD of the zeroth order, i.e.,  $ A(s_j) $. Next, using an arbitrariness of the choice of $ \mathcal{U} $ and $ \mathcal{V} $ originating from the zero matrix $ O_r $, the top-left $ d\times d $ submatrix of the first-order coefficient matrix can be also transformed into an SVD form, which is $ \tilde{\Lambda} $. Thus we obtain (\ref{eq:Assvd}). \\
	\indent  If we assume the invertibility of $ \tilde{\Lambda} $, which will be justified later in footnote \ref{foot1}, the expansion of  $ A(s)^{-1} $ is given by 
	\begin{align}
		A(s)^{-1}=\mathcal{V}\left[ \frac{1}{s-s_j}\begin{pmatrix} \tilde{\Lambda}^{-1} & \\ & O_{d-r} \end{pmatrix}+\begin{pmatrix} * & * \\ * & \Lambda^{-1}  \end{pmatrix} \right]\mathcal{U}^\dagger+O((s-s_j)^1), 
	\end{align}
	Integrating this expression on the contour $ |s-s_j|=\delta $ yields 
	\begin{align}
		C_j=\mathcal{V}\begin{pmatrix} \tilde{\Lambda}^{-1} & \\ & O_{d-r} \end{pmatrix} \mathcal{U}^\dagger, \label{eq:Cj001}
	\end{align}
	which obviously satisfies
		\begin{gather}
		A(s_j)C_j=C_jA(s_j)=0,\\
		\operatorname{rank}C_j=r.
	\end{gather}
	Since $ \operatorname{rank}C_j=r $, the column vectors in $ C_j $ span the null space of $ A(s_j) $, namely, if we write $ C_j=(c_{j1},\dots,c_{jd}) $, then $ \operatorname{span}\{c_{j1},\dots,c_{jd}\}=\operatorname{Ker}A(s_j) $.\\
	\indent Here we derive a few expressions necessary in the next subsection to calculate a normalization coefficient of bound states. Multiplying Eq.~(\ref{eq:scatinv2}) by  $ C_j^\dagger $ from the left and substituting $ s=s_j^*(=s_j^{-1}) $ yields $ C_j^\dagger B(s_j)^\dagger\Delta_-^\dagger A(s_j)=0 $, implying that 
	$ C_j^\dagger B(s_j)^\dagger\Delta_-^\dagger $ has the form
	\begin{align}
		C_j^\dagger B(s_j)^\dagger\Delta_-^\dagger=\mathcal{U}\begin{pmatrix} \tilde{B} & \\ & O_{d-r} \end{pmatrix} \mathcal{U}^\dagger, \label{eq:Cj4}
	\end{align}
	where $ \tilde{B} $ is an $ r\times r $ matrix and $ O_{d-r} $ is a zero matrix of size $ d-r $. Using Eqs.~(\ref{eq:Assvd}) and (\ref{eq:Cj001}),
	\begin{align}
		\dot{A}(s_j)C_j=\mathcal{U} \begin{pmatrix} I_r &  \\ * & O_{d-r} \end{pmatrix} \mathcal{U}^\dagger. \label{eq:Cj005}
	\end{align}
	Combining Eqs. (\ref{eq:Cj4}) and (\ref{eq:Cj005}) gives
	\begin{align}
		C_j^\dagger B(s_j)^\dagger \Delta_-^\dagger \dot{A}(s_j)C_j=C_j^\dagger B(s_j)^\dagger \Delta_-^\dagger. \label{eq:Cj6}
	\end{align}
\subsubsection{$ H_j(x) $: the orthonormalized bound states}\label{subsubsec:Hj}
	\indent The right and left Jost functions are related as [Eq.~(\ref{eq:defSmat})]
	\begin{align}
		F_+(x,s)\Delta_+^\dagger=F_-(x,s)\Delta_-^\dagger A(s)+s F_-(x,s^{-1})B(s) \label{eq:jostbound12}
	\end{align}
	Substituting $ s=s_j $ in this expression, and multiplying $ C_j $ from right yields:
	\begin{align}
		F_+(x,s_j)\Delta_+^\dagger C_j=s_j F_-(x,s_j^{-1})B(s_j)C_j=:-\frac{2\mathrm{i}s_j}{m}H_j(x)P_j^\dagger=-\frac{2\mathrm{i}s_j}{m}\sum_{i=1}^rh_{ji}(x)c_{ji}\hat{p}_{ji}^\dagger, \label{eq:jostbound11}
	\end{align}
	where we define rank-$r$  $ d\times r $ matrices $ H_j(x) $ and $ P_j $ by
	\begin{align}
		H_j(x)&=(h_{j1}(x),\dots,h_{jr}(x)),\quad \int_{-\infty}^\infty \mathrm{d}x H_j(x)^\dagger H_j(x)=I_r,  \label{eq:jostbound112} \\
		P_j&=(c_{j1}\hat{p}_{j1},\dots,c_{jr}\hat{p}_{jr}),\quad \hat{p}_{ji}^\dagger\hat{p}_{jl}=\delta_{il},\ c_{ji}>0. \label{eq:jostbound113}
	\end{align}
	$ h_{j1}(x),\dots,h_{jr}(x) $ are orthonormalized bound states of eigenvalue $ \epsilon(s_j) $. The factor $-\frac{2\mathrm{i}s_j}{m}$ is introduced just for later convenience.  $ P_j $ is a coefficient matrix of normalized bound states. Using an arbitrariness of the definition of $ H_j $ of unitary transformation $ H_j\rightarrow H_jU $, we can always choose $ P_j $ such that $ P_j^\dagger P_j $ is diagonal, and such $ P_j $ is already chosen in Eq. (\ref{eq:jostbound113}). In such a choice, the coefficient vectors becomes orthogonal $ \hat{p}_{ji}^\dagger\hat{p}_{jl}=\delta_{il} $ and the spectral decomposition of $ P_jP_j^\dagger $ is given by $ P_jP_j^\dagger=\sum_{i=1}^r c_{ji}^2\hat{p}_{ji}\hat{p}_{ji}^\dagger $. Using these notations, 
	\begin{align}
		\int_{-\infty}^\infty\mathrm{d}x \left(F_+(x,s_j)\Delta_+^\dagger C_j\right)^\dagger F_+(x,s_j)\Delta_+^\dagger C_j&=\frac{4}{m^2}\int_{-\infty}^\infty\mathrm{d}x P_j H_j^\dagger H_j P_j^\dagger=\frac{4}{m^2}P_jP_j^\dagger. \label{eq:boundintegral}
	\end{align}
	Let us derive another expression for this integral. The asymptotic forms of the Jost function and its  $ s $-derivative (denoted by dot) are given by 
	\begin{align}
		F_+(x,s_j)\Delta_+^\dagger C_j &\rightarrow \begin{cases} s_j\Phi_-(x,s_j^{-1})B(s_j)C_j &(x\rightarrow-\infty) \\ \Phi_+(x,s_j)\Delta_+^\dagger C_j &(x\rightarrow+\infty), \end{cases} \label{eq:Cj7} \\
		\dot{F}_+(x,s_j)\Delta_+^\dagger C_j &\rightarrow \begin{cases} \Phi_-(x,s_j)\dot{A}(s_j)C_j+ \frac{\mathrm{d} }{\mathrm{d} s}[s\Phi_-(x,s^{-1})B(s)]_{s=s_j}C_j &(x\rightarrow-\infty) \\ \dot{\Phi}_+(x,s_j)\Delta_+^\dagger C_j &(x\rightarrow+\infty), \end{cases} \label{eq:Cj75}
	\end{align}
	Using Eqs. (\ref{eq:boundprod}), (\ref{eq:constsolJ}), (\ref{eq:Cj6}), (\ref{eq:Cj7}), and (\ref{eq:Cj75}), we obtain
	\begin{align}
		\int_{-\infty}^\infty\mathrm{d}x \left(F_+(x,s_j)\Delta_+^\dagger C_j\right)^\dagger F_+(x,s_j)\Delta_+^\dagger C_j&=\frac{-\mathrm{i}}{\dot{\epsilon}(s_j)}\left[ \left( F_+(x,s_j)\Delta_+^\dagger C_j \right)^\dagger\sigma_3\left( \dot{F}_+(x,s_j)\Delta_+^\dagger C_j \right) \right]_{-\infty}^\infty \nonumber \\
		&=\frac{2\mathrm{i}s_j}{m}C_j^\dagger B(s_j)^\dagger \Delta_-^\dagger. \label{eq:boundintegral2}
	\end{align}
	Thus, from Eqs.~(\ref{eq:boundintegral}) and (\ref{eq:boundintegral2}),
	\begin{align}
		\frac{2\mathrm{i}s_j}{m}\Delta_-^\dagger P_jP_j^\dagger=B(s_j)C_j \label{eq:PPdaggerBC}
	\end{align}
	Since $ P_j $ has rank $ r $, there is a matrix $ X $ such that $ P_jX=I_r $. Therefore, Eqs.~(\ref{eq:jostbound11}) and (\ref{eq:PPdaggerBC}) imply
	\begin{align}
		H_j(x)=-s_jF_-(x,s_j^{-1})\Delta_-^\dagger P_j, \label{eq:boundrel12}
	\end{align}
	which expresses normalized bound states by scattering data.\footnote{\label{foot1}
	Here, we provide a proof that the diagonal elements of $ \tilde{\Lambda} $ in Eq.~(\ref{eq:Assvd}) are all positive. As an example, let us assume $ \tilde{\Lambda}_{11}=0 $. We modify the definition of $ C_j $ in Eq. (\ref{eq:Cj001}) by replacing the first diagonal element by $ 1 $. Then, $ A(s_j)C_j=C_jA(s_j)=0 $ and $ \operatorname{rank} C_j=r $, and hence this modified $ C_j $ can also extract all $ r $ independent bound states. Equation (\ref{eq:Cj4}) is also unchanged. Note that $ \tilde{B} $ in Eq.~(\ref{eq:Cj4}) must have rank $ r $, since the asymptotic forms of Eq. (\ref{eq:Cj7}) at $ x=+\infty $ and $ x=-\infty $ must have the same rank (because there must exist the same number of linearly independent solutions for given $r$ linearly independent initial conditions.). On the other hand, in Eq. (\ref{eq:Cj005}),  $ I_r $ should be replaced by $ 0\oplus I_{r-1} $ since $ \tilde{\Lambda}_{11}=0 $. Therefore, Eq. (\ref{eq:Cj6}) does not hold and we conclude that $ \operatorname{rank} [C_j^\dagger B(s_j)^\dagger\Delta_-^\dagger \dot{A}(s_j)C_j]=r-1 < \operatorname{rank} [C_j^\dagger B(s_j)^\dagger\Delta_-^\dagger]=r $. Thus the RHS of Eq. (\ref{eq:boundintegral2}) is replaced by $ \frac{2\mathrm{i}s_j}{m}C_j^\dagger B(s_j)^\dagger\Delta_-^\dagger \dot{A}(s_j)C_j $ and this expression has rank $ r-1 $, while Eq. (\ref{eq:boundintegral}) has rank $ r $, giving a contradiction. 
	} \\ 
\subsubsection{Symmetric and antisymmetric cases}\label{subsec:boundsymantisym}
	\indent Finally, we derive a constraint imposed on $ P_j $ when $ \Delta(x) $ is symmetric or antisymmetric. 
	The relation $ R(s)=\pm R(s)^T $ holds when $ \Delta(x)=\pm\Delta(x)^T $ [Eqs. (\ref{eq:absym}) and (\ref{eq:abasym})]. Integrating it on the contour $ |s-s_j|=\delta $ yields
	\begin{align}
		B(s_j)C_j=\pm(B(s_j)C_j)^T, \quad (\text{for }\Delta(x)=\pm\Delta(x)^T).
	\end{align}
	Thus, $ \Delta_-^\dagger \tilde{P}_j\tilde{P}_j^\dagger $ becomes a symmetric or antisymmetric matrix. If $ \Delta(x)=\pm\Delta(x)^T $, it also holds for $ x=-\infty $, and hence $ \Delta_-=\pm\Delta_-^T $. Therefore, the relation
	\begin{align}
		(P_jP_j^\dagger)^*=\Delta_-^\dagger(P_jP_j^\dagger)\Delta_-
	\end{align}
	holds for both symmetric and antisymmetric cases. From this, we soon find that if $ \hat{p}_{ji} $ is an eigenvector with eigenvalue $ c_{ji}^2 $, then $ \Delta_-\hat{p}_{ji}^* $ is also an eigenvector with the same eigenvalue:
	\begin{align}
		P_jP_j^\dagger \hat{p}_{ji}=c_{ji}^2\hat{p}_{ji} \quad\leftrightarrow\quad P_jP_j^\dagger \Delta_- \hat{p}_{ji}^*=c_{ji}^2 \Delta_-\hat{p}_{ji}^*.
	\end{align}
	We can verify that $ \hat{p}_{ji} $ and $ \Delta_-\hat{p}_{ji}^* $ are linearly dependent if $ \Delta_-=\Delta_-^T $, while linearly independent if $ \Delta_-=-\Delta_-^T $. Therefore, when $ \Delta(x) $ is symmetric,  $ \Delta_-^{-1/2}\hat{p}_{ji} $ can be chosen to be a real vector. On the other hand, if $ \Delta(x) $ is antisymmetric, two eigenvectors $ \hat{p}_{ji} $ and $ \Delta_-\hat{p}_{ji}^* $ always emerge in pairs. Thus, the degeneracy of eigenvalues of bound states is always even. 

\subsection{Integral representation of Jost function}\label{sec:kernelK}
	\indent Now we introduce an integral representation for left Jost functions using the kernel $ K(x,y) $: 
	\begin{align}
		F_-(x,s)&=\Phi_-(x,s)+\int_{-\infty}^x\mathrm{d}yK(x,y)\Phi_-(y,s), \label{eq:JostkernelKl0}
	\end{align}
	where $ K(x,y) $ is assumed to decrease exponentially for $ y\rightarrow-\infty $. This representation is allowed for $ s $ such that the integrand does not diverge at $ y=-\infty $. Such a region at least includes $ \operatorname{Im}s \le 0 $ as a subset. Replacing $ s\rightarrow s^{-1} $, we obtain another representation
	\begin{align}
		sF_-(x,s^{-1})=s\Phi_-(x,s^{-1})+\int_{-\infty}^x\mathrm{d}yK(x,y)s\Phi_-(y,s^{-1}), \label{eq:JostkernelKr0}
	\end{align}
	which is valid at least for $\operatorname{Im}s \ge 0  $. Equations~(\ref{eq:JostkernelKl0}) and (\ref{eq:JostkernelKr0}) are collectively written as
	\begin{align}
		Y_-(x,s)=\Psi_-(x,s)+\int_{-\infty}^x\mathrm{d}yK(x,y)\Psi_-(y,s). \label{eq:JostkernelK}
	\end{align}
	However, we should keep in mind that the right $ d $ columns and the left $ d $ columns in Eq. (\ref{eq:JostkernelK}) have different regions where the integral representation is allowed. When $ s \in \mathbb{R} $, all columns allow the integral representation. \\ 
	\indent The kernel $ K $ satisfies two important equations. To write down this, we write the matrix ZS operator for uniform and non-uniform $\Delta$ as
	\begin{align}
		\mathcal{L}(x)=\begin{pmatrix} -\mathrm{i}\partial_xI_d & \Delta(x) \\ \Delta(x)^\dagger & \mathrm{i}\partial_x I_d \end{pmatrix},\quad \mathcal{L}_0(x)=\begin{pmatrix} -\mathrm{i}\partial_xI_d & m\Delta_- \\ m\Delta_-^\dagger & \mathrm{i}\partial_x I_d \end{pmatrix}=-\mathrm{i}\sigma_3\partial_x+mM_-.
	\end{align}
	We also define
	\begin{align}
		\mathcal{V}(x):=\mathcal{L}(x)-\mathcal{L}_0(x), 
	\end{align}
	Then, the kernel $ K(x,y) $ satisfies
	\begin{align}
		\mathrm{i}[\sigma_3,K(x,x)]&=\mathcal{V}(x), \label{eq:kernelKdiff21} \\  
		\mathcal{L}(x)K(x,y)&=K(x,y)\mathcal{L}^{(0)}(y). \label{eq:kernelKdiff2}
	\end{align}
	Here, the operation of $ \mathcal{L}^{(0)} $ from right is defined by $ f\mathcal{L}_0(y):=\mathrm{i}\partial_yf\sigma_3+fmM_- $.   
	
	The derivation is as follows. Note that $ Y_- $ and $ \Psi_- $ satisfy $ \mathcal{L}Y_-=\epsilon Y_- $ and $ \mathcal{L}_0\Psi_-=\epsilon\Psi_- $.  Operating $ \mathcal{L}(x) $ to Eq. (\ref{eq:JostkernelK}) yields $ \mathcal{L}(x)Y_-(x,s)=\mathcal{V}(x)\Psi_-(x,s)+\epsilon\Psi_-(x,s)-\mathrm{i}\sigma_3 K(x,x)\Psi_-(x,s)+\int\mathrm{d}y[\mathcal{L}(x)K(x,y)]\Psi_-(y,s) $. Multiplying $ \epsilon $ to Eq. (\ref{eq:JostkernelK}) and integration by part yields $ \epsilon Y_-(x,s)=\epsilon\Psi_-(x,s)-\mathrm{i}K(x,x)\sigma_3\Psi_-(x,s)+\int\mathrm{d}y[K(x,y)\mathcal{L}_0(y)]\Psi_-(y,s) $. Equating these two, which holds for all $ s\in\mathbb{R} $, yields (\ref{eq:kernelKdiff21}) and (\ref{eq:kernelKdiff2}). \\
	\indent When $ \Delta(x) $ is symmetric or antisymmetric, the complex conjugation relations are given by $ \sigma_1 K(x,y)^*\sigma_1=K(x,y) $ or $ \sigma_2 K(x,y)^*\sigma_2=K(x,y) $, respectively. \\
	\indent The existence of the solution of the differential equation (\ref{eq:kernelKdiff2}), which guarantees the use of the integral representation (\ref{eq:JostkernelKl0}), is shown in Appendix~\ref{app:Kintegral}. \\ 
	\indent Equation (\ref{eq:kernelKdiff21}) is used to express $ \Delta $ by $ K $: 
	\begin{align}
		\Delta(x)=m\Delta_-+2\mathrm{i}K_{12}(x,x),\quad \Delta(x)^\dagger=m\Delta_-^\dagger-2\mathrm{i}K_{21}(x,x). \label{eq:kernelKdiff215}
	\end{align}
	Equation (\ref{eq:kernelKdiff2}) implies that the kernel $ K(x,y) $ must have the following expression:
	\begin{align}
		K(x,y) \sim \sum_\epsilon f(x,\epsilon)\phi(y,\epsilon^*)^\dagger, \label{eq:kernelKdiff4}
	\end{align}
	where $ f(x,\epsilon) $ is an eigenstate of $ \mathcal{L}(x) $ with eigenvalue $ \epsilon $, and $ \phi(y,\epsilon^*) $ is an eigenstate of $ \mathcal{L}^{(0)}(y) $ with eigenvalue $ \epsilon^* $, and the summation for $ \epsilon $ may include both discrete and continuous one (i.e., an integral).  The boundedness of $ K(x,y) $ with respect to $ x $ in fact suggests that the range of summation is restricted to $ \epsilon \in \mathbb{R} $. The expression of $ K $ in the form (\ref{eq:kernelKdiff4}) will be given in Subsec.~\ref{eq:kerneltojost}.
	
\subsection{Gelfand-Levitan-Marchenko equation}\label{subsec:glm}
	In this subsection, we derive the GLM equation. From the integral representation introduced above and Eq.~(\ref{eq:jostbound12}),
	\begin{align}
		&F_+(x,s)\Delta_+^\dagger A(s)^{-1}-\Phi_-(x,s)\Delta_-^\dagger=\int_{-\infty}^x\mathrm{d}zK(x,z)\Phi_-(z,s)\Delta_-^\dagger\nonumber\\
		&\qquad\qquad\qquad+\left[ s\Phi_-(x,s^{-1})+\int_{-\infty}^x\mathrm{d}zK(x,z)s\Phi_-(z,s^{-1}) \right]R(s), \label{eq:deriveglm001}
	\end{align}
	where $ R(s)=B(s)A(s)^{-1} $ is the reflection coefficient matrix. 
	To derive the GLM equation, we calculate
	\begin{align}
		\frac{m}{2}\int_{-\infty}^{\infty}\frac{\mathrm{d}s}{s^2} \text{[Eq. (\ref{eq:deriveglm001})]}\left[ \Phi_-(y,s^*)\Delta_-^\dagger \right]^\dagger \quad (y<x), \label{eq:deriveGLM}
	\end{align}
	First, we evaluate the right-hand side. The completeness relation in the uniform system $ \Delta(x)=m\Delta_- $ is given by
	\begin{align}
		\frac{m}{2}\int_{-\infty}^{\infty}\frac{\mathrm{d}s}{s^2}\left( \Phi_-(z,s)\Delta_-^\dagger \right)\left( \Phi_-(y,s)\Delta_-^\dagger \right)^\dagger=\frac{m}{2}\int_{-\infty}^{\infty}\frac{\mathrm{d}s}{s^2}\begin{pmatrix} s^2I_d & s\Delta_- \\ s\Delta_-^\dagger &I_d \end{pmatrix}\mathrm{e}^{\mathrm{i}k(s)(z-y)}=2\pi \delta(z-y) I_{2d}. \label{eq:complenetess00}
	\end{align}
	Using this, if we define
	\begin{align}
		\Omega_{\text{sc}}(x,y):=\frac{m}{4\pi}\int_{-\infty}^{\infty}\frac{\mathrm{d}s}{s^2}\left[s \Phi_-(x,s^{-1})\right]R(s)\left[ \Phi_-(y,s^*)\Delta_-^\dagger \right]^\dagger, \label{eq:omegas}
	\end{align}
	then the right-hand side is summarized as
	\begin{align}
		\text{[R.H.S of Eq.(\ref{eq:deriveGLM})]}=2\pi K(x,y)+2\pi \Omega_{\text{sc}}(x,y)+2\pi \int_{-\infty}^x\mathrm{d}z K(x,z)\Omega_{\text{sc}}(z,y).
	\end{align}
	Next, we evaluate the left-hand side. 
	The integrand of left-hand side of Eq.~(\ref{eq:deriveGLM}) is written only by $ s $ and does not include $ s^* $, and hence it is meromorphic about $ s $ and the residue theorem is applicable.
	We consider the semicircular contour in the upper half-plane (Fig.~\ref{fig:contours}). The contribution from the arc vanishes in the infinite-radius limit, and we obtain 
	\begin{align}
		\text{[L.H.S of Eq.(\ref{eq:deriveGLM})]}&=2\pi\mathrm{i}\frac{m}{2}\sum_j s_j^{-2}\left[F_+(x,s_j)\Delta_+^\dagger C_j\right] \left[\Phi_-(y,s_j^*)\Delta_-^\dagger\right]^\dagger \nonumber \\
		&=-2\pi\sum_j s_j F_-(x,s_j^{-1})\Delta_-^\dagger P_jP_j^\dagger \left[s_j\Phi_-(y,s_j^{-1})\Delta_-^\dagger\right]^\dagger,
	\end{align}
	where Eqs.~(\ref{eq:DefCj}), (\ref{eq:jostbound11}), (\ref{eq:PPdaggerBC}) and the relation $ s_j^{-1}=s_j^* $ are used.
	If we introduce
	\begin{align}
		\Omega_{\text{bd}}(x,y):=\sum_j \left[s_j\Phi_-(x,s_j^{-1})\Delta_-^\dagger\right] P_jP_j^\dagger \left[s_j\Phi_-(y,s_j^{-1})\Delta_-^\dagger\right]^\dagger, \label{eq:defomegad}
	\end{align}
	then, using Eq.~(\ref{eq:JostkernelKr0}), the left-hand side becomes
	\begin{align}
		\text{[L.H.S of Eq.(\ref{eq:deriveGLM})]}=-2\pi \Omega_{\text{bd}}(x,y)-2\pi \int_{-\infty}^x\mathrm{d}zK(x,z)\Omega_{\text{bd}}(z,y).
	\end{align}
	Summarizing, we obtain the GLM equation given as follows:
	\begin{align}
		K(x,y)+\Omega(x,y)+\int_{-\infty}^x\mathrm{d}zK(x,z)\Omega(z,y)=0, \label{eq:GLM00} \\
		\Omega(x,y)=\Omega_{\text{sc}}(x,y)+\Omega_{\text{bd}}(x,y).
	\end{align}
	The subscripts ``sc'' and ``bd'' represent the contributions from scattering and bound states, respectively. 
	The equation is derived under the assumption $ y<x $, while we need the case $ y=x $ when we calculate the potential [Eq.~(\ref{eq:kernelKdiff215})]. It should be interpreted as a limit $ y\rightarrow x-0 $. \\
	\indent Henceforth, we relabel the discrete eigenvalues by distinguishing all multiple zeros, and the corresponding bound states are also relabeled in the same manner. For example, if there are triply degenerate $ s_1 $, simple $ s_2 $, and doubly degenerate $ s_3 $, we relabel them as follows:
	\begin{align}
		(s_1,s_1,s_1,s_2,s_3,s_3)\quad\rightarrow\quad (s_1,s_2,s_3,s_4,s_5,s_6),
	\end{align} 
	and, correspondingly, we relabel the bound states and coefficient vectors introduced in Eqs. (\ref{eq:jostbound11})-(\ref{eq:jostbound113}) as
	\begin{align}
		(h_{11},h_{12},h_{13},h_{21},h_{31},h_{32}) &\quad\rightarrow\quad (h_1,h_2,h_3,h_4,h_5,h_6), \label{eq:relabelbound02} \\
		(c_{11}\hat{p}_{11},c_{12}\hat{p}_{12},c_{13}\hat{p}_{13},c_{21}\hat{p}_{21},c_{31}\hat{p}_{31},c_{32}\hat{p}_{32}) &\quad\rightarrow\quad (c_1\hat{p}_1,c_2\hat{p}_2,c_3\hat{p}_3,c_4\hat{p}_4,c_5\hat{p}_5,c_6\hat{p}_6).
	\end{align}
	In this new labeling,  $ \Omega_{\text{bd}}(x,y) $ [Eq.~(\ref{eq:defomegad})] is rewritten as
	\begin{align}
		\Omega_{\text{bd}}(x,y)&=W(x)W(y)^\dagger, \label{eq:omegab2} \\
		W(x)&=\left( s_1\Phi_-(x,s_1^{-1})\Delta_-^\dagger c_1\hat{p}_1,\dots,s_n\Phi_-(x,s_n^{-1})\Delta_-^\dagger c_n\hat{p}_n \right), \label{eq:omegab3}
	\end{align}
	where  $ n $ is a total number of bound states, and $ \hat{p}_i $'s satisfy 
	\begin{align}
		\hat{p}_i^\dagger\hat{p}_j=\begin{cases} \delta_{ij} & (\text{if } s_i=s_j), \\ \text{no constraint} & (\text{if } s_i\ne s_j). \end{cases}
	\end{align}
	We emphasize that $ \hat{p}_i=\hat{p}_j $ is possible if they do not belong to the same eigenvalue. Since $ \hat{p}_i $ is $ d $-component, the possible maximum degeneracy of discrete eigenvalues is $ d $. \\
	\indent In this labeling, the symmetric and antisymmetric conditions described in Subsec.~\ref{subsec:boundsymantisym} are rewritten as follows:
	\begin{gather}
		\hat{p}_j=\Delta_-\hat{p}_j^* \quad (\text{for the symmetric case}), \\
		\hat{p}_{2j}=\Delta_-\hat{p}_{2j-1}^*,\quad c_{2j}=c_{2j-1},\quad s_{2j}=s_{2j-1} \quad (\text{for the antisymmetric case}),
	\end{gather}
	where the pair of eigenstates for antisymmetric case is labeled by $ 2j-1 $ and $ 2j $. \\
	\indent The existence and the uniqueness of the solution of the GLM equation (\ref{eq:GLM00}) is proved in Appendix~\ref{app:glmuniq}. 
	An analysis of the reflectionless solution is given in Sec.~\ref{sec:asymptotics}.

\subsection{Kernel $ K(x,y) $ expressed by Jost functions}\label{eq:kerneltojost}
	As discussed in Subsec.~\ref{sec:kernelK}, the kernel $ K(x,y) $ should have the form of Eq.~(\ref{eq:kernelKdiff4}). The integral representation (\ref{eq:JostkernelKl0}) and the GLM equation (\ref{eq:GLM00}) imply that such an expression is given by
	\begin{align}
		K(x,y)&=K_{\text{sc}}(x,y)+K_{\text{bd}}(x,y), \label{eq:Kbyjost06} \\
		K_{\text{sc}}(x,y)&:=\frac{m}{4\pi}\int_{-\infty}^{\infty}\frac{\mathrm{d}\zeta}{\zeta^2}\left[ -\zeta F_-(x,\zeta^{-1}) \right]R(\zeta)\left[ \Phi_-(y,\zeta^*)\Delta^\dagger \right]^\dagger, \label{eq:Kbyjost07}\\
		K_{\text{bd}}(x,y)&:=H(x)W(y)^\dagger, \label{eq:Kbyjost08} \\
		H(x)&=(h_1(x),\dots,h_n(x))=-\left(s_1F_-(x,s_1^{-1})\Delta_-^\dagger c_1\hat{p}_1,\dots,s_nF_-(x,s_n^{-1})\Delta_-^\dagger c_n\hat{p}_n\right). \label{eq:Kbyjost09}
	\end{align}
	Here, $ H(x) $ is an array of normalized bound states, as derived in Subsec.~\ref{subsec:boundstates} and relabeled as Eq. (\ref{eq:relabelbound02}).  
	Substitution of these expressions into the GLM equation (\ref{eq:GLM00}) again reproduces the integral representation (\ref{eq:JostkernelKl0}). Note that $ K_{\text{sc}} $ and $ K_{\text{bd}} $ are \textit{formally} obtained by replacing $ \Phi_- \rightarrow -F_- $ in  $ \Omega_{\text{sc}} $ and $ \Omega_{\text{bd}} $, which are  defined by Eqs. (\ref{eq:omegas}), (\ref{eq:omegab2}), and (\ref{eq:omegab3}). 
	Because of the uniqueness of the solution of the GLM equation (Appendix~\ref{app:glmuniq}), if we find a solution, then this is the unique solution. Therefore $ K $ must be given by (\ref{eq:Kbyjost06})-(\ref{eq:Kbyjost09}). \\
	\indent We must carefully change the order of the $ \zeta $-integral and the $ y $-integral when we substitute the above kernel into the integral representation of the Jost function [Eqs.~(\ref{eq:JostkernelKl0}) and (\ref{eq:JostkernelKr0})], since the integrand is not $ L^1 $-integrable. In order to apply the Fubini-Tonelli theorem, we must first consider $ \int_{-\infty+\mathrm{i}\delta}^{\infty+\mathrm{i}\delta}\mathrm{d}\zeta $, and take the limit $ \delta\rightarrow+0 $ after changing the order of the integral. In order to perform such calculations quickly, it is convenient to always add $ +\mathrm{i}0 $ in Eq. (\ref{eq:Kbyjost07}):
	\begin{align}
		K_{\text{sc}}(x,y)&=\frac{m}{4\pi}\int_{-\infty+\mathrm{i}0}^{\infty+\mathrm{i}0}\frac{\mathrm{d}\zeta}{\zeta^2}\left[ -\zeta F_-(x,\zeta^{-1}) \right]R(\zeta)\left[ \Phi_-(y,\zeta^*)\Delta^\dagger \right]^\dagger.
	\end{align}
	The change of the variable $ \zeta\rightarrow\zeta^{-1} $ yields another expression:
	\begin{align}
		K_{\text{sc}}(x,y)&=\frac{m}{4\pi}\int_{-\infty-\mathrm{i}0}^{\infty-\mathrm{i}0}\frac{\mathrm{d}\zeta}{\zeta^2}\left[ - F_-(x,\zeta)\Delta^\dagger \right]R(\zeta^*)^\dagger \left[ \zeta^*\Phi_-(y,\zeta^{-1*}) \right]^\dagger.
	\end{align}
	
	\indent Substituting $ s=s_j $ in Eq.~(\ref{eq:JostkernelKr0}), multiplying $ \Delta_-^\dagger c_j\hat{p}_j $ from right, and making an array by collecting them for $ j=1,\dots, n $ yields
	\begin{align}
		-H(x)=W(x)+\int_{-\infty}^x\mathrm{d}yK(x,y)W(y).
	\end{align}
	If $ K_{\text{sc}}(x,y)=0 $, this is the same as the ``GLM ansatz'' in Ref.~\cite{arxiv1509.04242}. 
	
	Let us define
	\begin{align}
		G(x)&:=\int_{-\infty}^x\mathrm{d}y W(y)^\dagger W(y), \label{eq:Ktojostker1} \\
		\Gamma_-(x,s)&:=\int_{-\infty}^x\mathrm{d}yW(y)^\dagger \Phi_-(y,s),  \label{eq:Ktojostker2} \\
		X(x,\zeta,s)&:=\int_{-\infty}^x\mathrm{d}y \left[ \Phi_-(y,\zeta^*)\Delta^\dagger \right]^\dagger\Phi_-(y,s), \label{eq:Ktojostker3} 
	\end{align}
	where, in the last expression, at least one of $ \zeta $ or $ s^{-1} $ must have $ +\mathrm{i}0 $ to ensure the convergence at $ x=-\infty $. 
	Then, the left Jost functions are rewritten as 
	\begin{align}
		-H(x)&=W(x)+H(x)G(x)+\frac{m}{4\pi}\int_{-\infty+\mathrm{i}0}^{\infty+\mathrm{i0}}\frac{\mathrm{d}\zeta}{\zeta^2}\left[ -\zeta F_-(x,\zeta^{-1}) \right]R(\zeta)[\Gamma_-(x,\zeta^*)\Delta_-^\dagger]^\dagger, \label{eq:HjostRH} \\
		F_-(x,s)&=\Phi_-(x,s)+H(x)\Gamma_-(x,s)+\frac{m}{4\pi}\int_{-\infty+\mathrm{i}0}^{\infty+\mathrm{i0}}\frac{\mathrm{d}\zeta}{\zeta^2}\left[ -\zeta F_-(x,\zeta^{-1}) \right]R(\zeta) X(x,\zeta,s). 
	\end{align}
	These equations are probably the same as the ones that are directly derived by the Riemann-Hilbert method \cite{FaddeevTakhtajan,AblowitzSegur,Ablowitzbook}). 
%

\subsection{Orthonormal basis of scattering eigenstates}\label{subsec:ortho}
	The Jost functions  $ Y_\pm(x,s) $ with $ s\in \mathbb{R},\ s\ne 1 $ generally form a linearly independent basis for a given eigenvalue $ \epsilon(s) $. However, the columns of $ Y_\pm(x,s) $ are not generally orthogonal to each other. Here, we construct an orthonormal basis. Let us define 
	\begin{align}
		Q(s):=\left(\frac{I_{2d}+S(s)^\dagger S(s)}{2} \right)^{-1/2},\quad s\in\mathbb{R}.
	\end{align}
	Since $ I_{2d}+S(s)^\dagger S(s) $ is a positive-definite hermitian matrix, its square root inverse is defined unambiguously. 
	 $ Q(s)=I_{2d} $ for a reflectionless potential. 
	$ Q(s) $ itself does not have a closed form, but its square can be written as
	\begin{align}
		Q(s)^2=\begin{pmatrix} I_d & -A(s)^{-1}\Delta_- B(s^{-1})\Delta_+ \\ -\Delta_+^\dagger A(s^{-1})^{-1}\Delta_- B(s) & I_d \end{pmatrix},
	\end{align}
	whose hermiticity is checked by using Eq. (\ref{eq:scatinv42}).  $ Q(s) $ satisfies the following properties:
	\begin{align}
		& Q(s^{-1})=M_+Q(s)M_+, \label{eq:orthoQpr1} \\
		&Q(s)S(s)^\dagger S(s)Q(s)=\sigma_3 Q(s)^2\sigma_3, \label{eq:QSSQesQ2s} \\
		&S(s)Q(s)^2S(s)^\dagger=\begin{pmatrix} I_d & R(s)^\dagger \\ R(s) & I_d \end{pmatrix}. \label{eq:SQQSeIRRI}
	\end{align}
	We define $ Y_o(x,s) $ and $ F_o(x,s) $ by
	\begin{align}
		Y_o(x,s):=\left( F_o(x,s)\Delta_+^\dagger, s F_o(x,s^{-1}) \right):=Y_+(x,s)Q(s)=Y_-(x,s)S(s)Q(s). \label{eq:YoFodef}
	\end{align}
	Here the subscript $ o $ means ``orthonormal''. \\
	\indent Let us show that the column vectors in $ Y_o(x,s) $ are orthonormal. 
	First, we consider a finite-size system in the interval $ [-L,L] $ and take $ L\rightarrow\infty $ later. Let us take $ s_1,s_2\in\mathbb{R} $ and write $ \epsilon_i=\epsilon(s_i),\ k_i=k(s_i) $. Then, using Eq.~(\ref{eq:constwjnew01}), and $ Y_o=Y_+Q $, 
	\begin{align}
		&\int_{-L}^L\mathrm{d}x Y_o(x,s_1)^\dagger Y_o(x,s_2)=\frac{[Q(s_1)Y_+(x,s_1)^\dagger\sigma_3 Y_+(x,s_2)Q(s_2)]_{-L}^L}{-\mathrm{i}(\epsilon_1-\epsilon_2)} 
	\end{align}
	If $ L $ is sufficiently large, the plane-wave asymptotic form (\ref{eq:asymptoticY}) can be used as a value at $ x=\pm L $. 
	The expression for $ \Psi_\pm $ is given by Eq.~(\ref{eq:Psiintroduce}). Using  $ \mathrm{e}^{\mathrm{i}kx\sigma_3}=(\cos kx)I_{2d}+(\mathrm{i}\sin kx)\sigma_3 $ and rewriting
	\begin{align}
		\frac{s_1s_2-1}{-\mathrm{i}(\epsilon_1-\epsilon_2)}=\frac{1+s_1s_2}{-\mathrm{i}(k_1-k_2)},\quad \frac{s_1-s_2}{-\mathrm{i}(\epsilon_1-\epsilon_2)}=\frac{s_1+s_2}{-\mathrm{i}(k_1+k_2)}, \label{eq:e1e2tok1k2}
	\end{align}
	and using the formulae $\lim_{L\rightarrow\infty}\mathrm{e}^{\mathrm{i}kL}=0  $ (as a distribution) and $ \lim_{L\rightarrow\infty}\frac{\sin Lk}{k}=\pi \delta(k) $, the limit $ L\rightarrow \infty $ yields
	\begin{align}
		\int_{-\infty}^\infty\mathrm{d}x Y_o(x,s_1)^\dagger Y_o(x,s_2)=2\pi(1+s_1s_2)\delta(k_1-k_2)I_{2d}+2\pi(s_1+s_2)\delta(k_1+k_2)M_+. \label{eq:ortho4}
	\end{align}
	If we extract the top-left $ d\times d $ block of Eq.~(\ref{eq:ortho4}),
	\begin{align}
		\int_{-\infty}^\infty\mathrm{d}x F_o(x,s_1)^\dagger F_o(x,s_2)=2\pi(1+s_1s_2)\delta(k_1-k_2)I_d.
	\end{align}
	Thus orthonormality is shown. 
	Here, in deriving Eq. (\ref{eq:ortho4}), we use the fact that the first (last) term of the right-hand side is nonvanishing only when $ s_1=s_2  \ (s_1=s_2^{-1}) $ and use $ 2Q^{-2}=I+S^\dagger S $ and Eq.~(\ref{eq:orthoQpr1}).

\subsection{Completeness relation}\label{subsec:ortho2}
	Let us derive the completeness relation of the Jost functions. Using the completeness relation of the uniform system (\ref{eq:complenetess00}) and the integral representation (\ref{eq:JostkernelKr0}), we obtain
	\begin{align}
		&\frac{m}{4\pi}\int_{-\infty}^\infty\frac{\mathrm{d}s}{s^2}\left[ sF_-(x,s^{-1}) \right]\left[ sF_-(y,s^{-1}) \right]^\dagger=\delta(x-y)I_{2d}\nonumber \\
		&+\theta(x-y)\left[ K(x,y)+\int_{-\infty}^y\mathrm{d}z K(x,z)K(y,z)^\dagger \right]+\theta(y-x)\left[ K(y,x)^\dagger+\int_{-\infty}^x\mathrm{d}z K(x,z)K(y,z)^\dagger \right]
	\end{align}
	Using the equations introduced in Subsec.~\ref{eq:kerneltojost} and the relation $ \Delta_-R(s^{-1})-R(s)^\dagger\Delta_-^\dagger=0 \ $ [Eq.~(\ref{eq:scatinv202}), $s\in\mathbb{R}$], we obtain
	\begin{align}
		&K(y,x)^\dagger+\int_{-\infty}^x\mathrm{d}z K(x,z)K(y,z)^\dagger=K(x,y)+\int_{-\infty}^y\mathrm{d}z K(x,z)K(y,z)^\dagger \nonumber \\
		&=-H(x)H(y)^\dagger-\frac{m}{4\pi}\int_{-\infty}^\infty\frac{\mathrm{d}s}{s^2}F_-(x,s)\Delta_-^\dagger R(s)^\dagger \left[ s F(y,s^{-1}) \right]^\dagger, \label{eq:completenessproof01}
	\end{align}
	where we omit $ +\mathrm{i}0 $ and write $ s^*=s $ since there is no convergence problem for this integral.
	Thus we get
	\begin{align}
		\frac{m}{4\pi}\int_{-\infty}^\infty\frac{\mathrm{d}s}{s^2}\left[ sF_-(x,s^{-1})+F_-(x,s)\Delta_-^\dagger R(s)^\dagger \right]\left[ sF_-(y,s^{-1}) \right]^\dagger+H(x)H(y)^\dagger =\delta(x-y)I_{2d}. \label{eq:completenesszero}
	\end{align}
	This provides a completeness relation, but it does not have a form like ``$ 1=\sum_n\ket{n}\bra{n} $''. Such an expression is obtained by rewriting it with $ F_o(x,s) $. 
	Using Eqs.~(\ref{eq:SQQSeIRRI}) and (\ref{eq:YoFodef}) and the relation $ \int_{-\infty}^\infty \frac{\mathrm{d}s}{s^2}F_\alpha(x,s)F_\alpha(y,s)^\dagger=\frac{1}{2}\int_{-\infty}^\infty \frac{\mathrm{d}s}{s^2}Y_\alpha(x,s)Y_\alpha(y,s)^\dagger \ (\alpha=o \text{ and }-) $, which is shown by the change of the integration variable $ s\rightarrow s^{-1} $, Eq.~(\ref{eq:completenesszero}) is rewritten as
	\begin{align}
		\frac{m}{4\pi}\int_{-\infty}^\infty \frac{\mathrm{d}s}{s^2}F_o(x,s)F_o(y,s)^\dagger+H(x)H(y)^\dagger=\delta(x-y)I_{2d}, \label{eq:completenessshin}
	\end{align}
	which is the completeness relation.

\section{Gap equation}\label{sec:gap}
	This section includes the main result of this paper. We introduce the matrix BdG and gap equations, and solve them.
\subsection{Gap equation in infinite-length system with finite reflection}
	\indent The BdG and the gap equations considered in Ref.~\cite{arxiv1509.04242} are as follows. The BdG equation is
	\begin{align}
		\epsilon \begin{pmatrix}\boldsymbol{u} \\ \boldsymbol{v}\end{pmatrix}=\begin{pmatrix}-\mathrm{i}\partial_x & \Delta(x) \\ \Delta(x)^\dagger & \mathrm{i}\partial_x \end{pmatrix}\begin{pmatrix}\boldsymbol{u} \\ \boldsymbol{v}\end{pmatrix}, \label{eq:gapintrobdg}
	\end{align}
	where $ \Delta(x) $ is a $ d\times d $ matrix and $ \boldsymbol{u},\boldsymbol{v} $ represent $ d $-component vectors. This is the same as the matrix ZS problem (\ref{eq:mZS}). As a self-consistent condition, three kinds of gap equations are considered; namely, the gap equations for symmetric and antisymmetric  $ \Delta $'s corresponding to $p$-wave and $s$-wave superfluids: 
	\begin{align}
		-\frac{\Delta}{g}=\frac{1}{2}\sum_j (\boldsymbol{u}_j\boldsymbol{v}_j^\dagger \pm \boldsymbol{v}_j^*\boldsymbol{u}_j^T)(2\nu_j-1) \quad\text{ for $\Delta=\pm\Delta^T$,}
	\end{align}
	and non-symmetric $ \Delta $ corresponding to $s$-$p$ mixed superfluids:
	\begin{align}
		-\frac{\Delta}{g}=\sum_j \boldsymbol{u}_j\boldsymbol{v}_j^\dagger(2\nu_j-1).
	\end{align}
	Here, $ \nu_j\in[0,1] $ denotes the filling rate for an eigenstate $ j $, and $ g>0 $ is a coupling constant. We use bold font $ (\boldsymbol{u},\boldsymbol{v}) $ for quasiparticle eigenfunctions  to avoid confusion with the filling $ \nu $. These equations appear when we consider a self-consistent solution in the BdG theory for unconventional and/or multicomponent fermionic condensates possessing $ SU(d) $-symmetric two-body interaction. The $s$-$p$ mixed superfluid is realized by fine tuning of coupling constants \cite{arxiv1509.04242}. Note that Eq.~(\ref{eq:gapintrobdg}) represents the BdG equation for right-mover eigenstates obtained by the Andreev approximation at $ k=k_F $. We also obtain left-mover ones linearized at $ k=-k_F $, but they need not be taken into account after elimination of double counting of the BdG eigenstates. (See Secs. IV and V of Ref.~\cite{arxiv1509.04242} for the Andreev approximation and the double-counting problem.) \\
	\indent Since we are interested in an excited state near the ground state, we consider the following occupation state: 
	\begin{align}
		\nu_j=\begin{cases} 1 & (\text{negative-energy scattering states}), \\ 0 & (\text{positive-energy scattering states}), \\ \text{adjust to solve the gap equation} & (\text{bound states}). \end{cases}
	\end{align}
	Let us define
	\begin{align}
		\Xi&=\Xi_{\text{eigenstates}}+\Xi_{\text{gap}},\\
		\Xi_{\text{eigenstates}}&=\sum_jw_jw_j^\dagger(2\nu_j-1)=\sum_j \begin{pmatrix} \boldsymbol{u}_j \\ \boldsymbol{v}_j \end{pmatrix}\begin{pmatrix} \boldsymbol{u}_j^\dagger & \boldsymbol{v}_j^\dagger \end{pmatrix}(2\nu_j-1),\\
		\Xi_{\text{gap}}&=\frac{1}{g}\begin{pmatrix} & \Delta \\ \Delta^\dagger & \end{pmatrix},
	\end{align}
	then the gap equation is equivalent to:
	\begin{align}
		\left[ \sigma_3, \Xi+\tau\Xi^*\tau \right]=0, \label{eq:gaptau001}
	\end{align}
	where $ \tau=0,\ \sigma_1, $ and $ \sigma_2 $ for non-symmetric, symmetric, and antisymmetric $ \Delta $, respectively. \\
	\indent We assume that $ \Delta $ has $ n $ bound states and write them as $ H=(h_1,\dots,h_n) $, and use the same notations for the scattering states as in the previous section. The discrete sum of scattering states in $ \Xi_{\text{eigenstates}} $ should be replaced by a continuous integral in the infinite-length limit. The orthonormal basis and its completeness relation derived in Subsecs. \ref{subsec:ortho} and \ref{subsec:ortho2} imply that the infinite-length limit is given by
	\begin{align}
		\Xi_{\text{eigenstates}}=HDH^\dagger+\left[\int_{-s_c}^{-s_c^{-1}}-\int_{s_c^{-1}}^{s_c}\right]\frac{m\mathrm{d}s}{4\pi s^2}F_o(x,s)F_o(x,s)^\dagger,
	\end{align}
	where $ D_{ij}=\delta_{ij}(2\nu_j-1) $ and  $ s_c $ is a cutoff, which is related to the momentum cutoff $ k_c $ by $ s_c=\frac{k_c+\sqrt{k_c^2+m^2}}{m}\simeq\frac{2k_c}{m} $. The cutoff $ k_c $ is determined by some other physical mechanism, e.g., the Debye frequency in condensed matter. \\
	\indent First, let us consider the uniform condensate $ \Delta(x)=m\Delta_- $. where scattering states are given by $ F_o(x,s)=\left( \begin{smallmatrix} s\Delta_- \\ I_d \end{smallmatrix} \right)\mathrm{e}^{\mathrm{i}k(s)x} $ and no bound state exists. The coupling constant $ g $, the cutoff $ k_c $, and the bulk gap $ m $  are related by 
	\begin{align}
		&-\frac{1}{g}=\left[\int_{s_c^{-1}}^{s_c}-\int_{-s_c}^{-s_c^{-1}}\right]\frac{\mathrm{d}s}{4\pi s} \\
		 \leftrightarrow\quad& s_c=\mathrm{e}^{\pi/g} \quad\leftrightarrow\quad m=2k_c\mathrm{e}^{-\pi/g}.
	\end{align}
	If we eliminate $ g $ using this relation, we can take the limit $ s_c\rightarrow+\infty $ since the logarithmically divergent terms in  $ \Xi_{\text{gap}} $ and $ \Xi_{\text{eigenstates}} $ cancel out. 
	The resultant expression is 
	\begin{align}
		\Xi=HDH^\dagger+\left[\int_{-\infty}^{0}-\int_{0}^{\infty}\right]\frac{m\mathrm{d}s}{4\pi}\left( \frac{F_o(x,s)F_o(x,s)^\dagger}{s^2}-\frac{1}{ms}\begin{pmatrix} & \Delta \\ \Delta^\dagger & \end{pmatrix} \right). \label{eq:gapxi002}
	\end{align}
	Note that the formal replacement $ \frac{1}{g}=\left[\int_{0}^{\infty}-\int_{-\infty}^{0}\right]\frac{\mathrm{d}s}{4\pi s}=\frac{1}{2\pi}\int_{-\infty}^\infty\frac{\mathrm{d}k}{\sqrt{k^2+m^2}} $ provides a shortcut to obtaining this expression. \\
	\indent Let us rewrite the gap equation using the left Jost function. For convenience, we introduce
	\begin{align}
		F(x,s):=s^{-1}F_-(x,s)\Delta_-^\dagger. \label{eq:gapFmvar}
	\end{align}
	If we restrict the problem to the reflectionless case, $ F(x,s)\mathrm{e}^{-\mathrm{i}\epsilon(s)t} $ corresponds to $ F(x,t,s) $ in Ref.~\cite{arxiv1509.04242}. The left Jost function  $ F_- $ and orthonormal Jost function $ F_o $ are related by Eq.~(\ref{eq:YoFodef}). Using Eqs.~(\ref{eq:scatinv202}), (\ref{eq:kernelKdiff21}), and (\ref{eq:SQQSeIRRI}),
	Eq. (\ref{eq:gapxi002}) is rewritten as
	\begin{align}
		\Xi&=HDH^\dagger\nonumber \\
		&\quad+\left[\int_{-\infty}^{0}-\int_{0}^{\infty}\right]\frac{m\mathrm{d}s}{4\pi}\left( F(x,s)F(x,s)^\dagger+F(x,s)R(s)^\dagger\Delta_-^\dagger\frac{F(x,s^{-1})^\dagger}{s}-\frac{mM_-+\mathrm{i}[\sigma_3,K(x,x)]}{ms} \right). \label{eq:gapxi001}
	\end{align}
	Thus, the gap equation that we must solve is  $ [\sigma_3,\Xi+\tau\Xi^*\tau]=0 $ with this $ \Xi $. 

\subsection{Proof of reflectionless nature}\label{subsec:solvegap}
	Let us solve the gap equation, Eq. (\ref{eq:gaptau001}) with (\ref{eq:gapxi001}) and (\ref{eq:gapFmvar}). The proof consists of the following three steps (I)-(III):
	\begin{enumerate}[(I) ]
		\item Rewrite $ \Xi $ as follows: $ \Xi=\Xi_{\text{bound}}+\Xi_{\text{scattering}} $ with
		\begin{align}
			\Xi_{\text{bound}}&=2H(\mathcal{N}-\Theta)H^\dagger,\quad \mathcal{N}=\operatorname{diag}(\nu_1,\dots,\nu_n),\ \Theta=\frac{1}{\pi}\operatorname{diag}(\theta_1,\dots,\theta_n),\quad \theta_j=\arg s_j. \label{eq:gapstep101}\\
			\Xi_{\text{scattering}}&=\int_{-\infty}^\infty\frac{m\mathrm{d}s}{8\pi s^2}Y_-(x,s)\mathcal{P}(s)Y_-(x,s)^\dagger,\quad \mathcal{P}(s)=\begin{pmatrix} & \frac{\log s^2}{\mathrm{i}\pi}R(s)^\dagger \\ -\frac{\log s^2}{\mathrm{i}\pi}R(s) &  \end{pmatrix}. \label{eq:gapstep102}
		\end{align}
		\item Using Eq.~(\ref{eq:constwjnew02}), calculate $ 0=\int\mathrm{d}x Y_+[\sigma_3,\Xi+\tau\Xi^*\tau]Y_+^\dagger $, and derive $ R(s)=0 $ (i.e.,  $ \Xi_{\text{scattering}}=0 $).
		\item Solve the bound-state part $ [\sigma_3,\Xi_{\text{bound}}+\tau\Xi_{\text{bound}}^*\tau]=0 $ and determine the condition for $ \nu_j $.   
	\end{enumerate}
	The order of (II) and (III) can be exchanged. Henceforth we perform the above (I)-(III). \\
	\indent \textit{\underline{Step (I):}} This step is nothing but a straightforward calculation, but since it is a little complicated, we show several key mathematical expressions.
	
	Let us write $ e_j(x)=c_j\mathrm{e}^{-\mathrm{i}k(s_j)x},\ j=1,\dots, n $. 
	Then $ W(x) $ defined by Eq.~(\ref{eq:omegab3}) is
	\begin{align}
		W=\begin{pmatrix} e_1\hat{p}_1 & \dots & e_n\hat{p}_n \\ s_1e_1 \Delta_-^\dagger\hat{p}_1 & \dots & s_ne_n\Delta_-^\dagger \hat{p}_n \end{pmatrix}.
	\end{align}
	Note that $ [e_j]_{\text{Ref.~\cite{arxiv1509.04242}}}\propto [e_j]_{\text{this paper}}\times \mathrm{e}^{-\mathrm{i}\epsilon(s_j)t} $, i.e., a time-dependent exponential factor is absent in the current treatment. $ G,\Gamma_-,X $ introduced in Eqs. (\ref{eq:Ktojostker1})-(\ref{eq:Ktojostker3}) are
	\begin{align}
		[G(x)]_{ij}&=-\frac{2\mathrm{i}}{m}\frac{s_i^{-1}s_j e_ie_j\hat{p}_i^\dagger\hat{p}_j}{s_i^{-1}-s_j}=\frac{2\mathrm{i}}{m}\frac{e_ie_j \hat{p}_i^\dagger\hat{p}_j}{s_i-s_j^{-1}},\label{eq:thegram} \\
		\Gamma_-(x,s)&=-\frac{2\mathrm{i}}{m}\begin{pmatrix} \frac{s}{s-s_1}e_1\hat{p}_1^\dagger\Delta_- \\ \vdots \\ \frac{s}{s-s_n}e_n\hat{p}_n^\dagger\Delta_- \end{pmatrix}\mathrm{e}^{\mathrm{i}k(s)x},\\
		s X(x,\zeta,s^{-1})&=\frac{(s+\zeta)\Delta_-}{-\mathrm{i}[k(\zeta)+k(s)]}\mathrm{e}^{-\mathrm{i}[k(s)+k(\zeta)]x},\quad (\zeta,s \in \mathbb{R}+\mathrm{i}0). 
	\end{align}
	Furthermore, we introduce
	\begin{align}
		\mathcal{R}(x,\zeta)=R(\zeta^*)^\dagger\Delta_-^\dagger\frac{F(x,\zeta^{-1*})^\dagger}{\zeta} \mathrm{e}^{\mathrm{i}k(\zeta)x},\quad \tilde{\mathcal{R}}(x,\zeta)=F(x,\zeta)R(\zeta^*)^\dagger\Delta_-^\dagger \mathrm{e}^{\mathrm{i}k(\zeta)x}. \label{eq:hilbertkernel}
	\end{align}
	For brevity, we henceforth abbreviate $ \int_{-\infty-\mathrm{i}0}^{\infty-\mathrm{i}0}\mathrm{d}\zeta $ as $ \int\mathrm{d}\zeta $. Then, the expressions of scattering and bound states are given by
	\begin{align}
		F(x,s)&=\left[ \begin{pmatrix} I_d \\ s^{-1}\Delta_-^\dagger \end{pmatrix}+\frac{2\mathrm{i}}{m}\sum_j\frac{h_je_j\hat{p}_j^\dagger}{s_j-s}+\int\frac{\mathrm{d}\zeta}{2\pi\mathrm{i}} \frac{\tilde{\mathcal{R}}(x,\zeta)}{1-s\zeta} \right]\mathrm{e}^{\mathrm{i}k(s)x}, \label{eq:pregap01} \\
		F(x,s^*)^\dagger&=\left[ \begin{pmatrix} I_d & s^{-1}\Delta_- \end{pmatrix}+\frac{2\mathrm{i}}{m}\sum_j\frac{e_j\hat{p}_jh_j^\dagger}{s-s_j^{-1}}+\int\frac{\mathrm{d}\zeta}{2\pi\mathrm{i}} \frac{\mathcal{R}(x,\zeta)}{s-\zeta} \right]\mathrm{e}^{-\mathrm{i}k(s)x}, \label{eq:pregap02} \\
		h_j(x)&=-\left[ \begin{pmatrix}I_d \\ s_j\Delta_-^\dagger \end{pmatrix}+\frac{2\mathrm{i}}{m}\sum_l\frac{h_le_l\hat{p}_l^\dagger}{s_l-s_j^{-1}}+\int\frac{\mathrm{d}\zeta}{2\pi\mathrm{i}}\frac{s_j\tilde{\mathcal{R}}(x,\zeta)}{s_j-\zeta} \right]e_j\hat{p}_j, \label{eq:pregap03} \\
		h_j(x)^\dagger&=-e_j\hat{p}_j^\dagger\left[ \begin{pmatrix} I_d & s_j^{-1}\Delta_- \end{pmatrix}+\frac{2\mathrm{i}}{m}\sum_l\frac{e_l\hat{p}_lh_l^\dagger}{s_j-s_l^{-1}}+\int\frac{\mathrm{d}\zeta}{2\pi\mathrm{i}}\frac{\mathcal{R}(x,\zeta)}{s_j-\zeta} \right]. \label{eq:pregap04}
	\end{align}
	Here and hereafter,  $ s $ and $ s^* $ in Eqs.~(\ref{eq:pregap01}) and (\ref{eq:pregap02}) should be interpreted as having an infinitesimal $ -\mathrm{i}0 $ and $ +\mathrm{i}0 $ if we encounter an integral not convergent for purely real $ s $. Otherwise, their difference is ignorable. 
	We also need an expression for the kernel $ K(x,x) $. Setting $ x=y $ in Eq.~(\ref{eq:completenessproof01}), we obtain $ K(x,x)^\dagger=K(x,x) $, i.e., it is self-adjoint. Using the self-adjointness,  two expressions are possible: $ K=K_{\text{sc}}+K_{\text{bd}}=K_{\text{sc}}^\dagger+K_{\text{bd}}^\dagger $ with 
	\begin{align}
		K_{\text{sc}}(x,x)&=-\frac{\mathrm{i}m}{2}\int\frac{\mathrm{d}\zeta}{2\pi\mathrm{i}} \tilde{\mathcal{R}}(x,\zeta)\begin{pmatrix} \zeta^{-1}I_d & \Delta_- \end{pmatrix}, \label{eq:pregap05} \\
		K_{\text{sc}}(x,x)^\dagger&=-\frac{\mathrm{i}m}{2}\int\frac{\mathrm{d}\zeta}{2\pi\mathrm{i}}\begin{pmatrix} I_d \\ \zeta^{-1}\Delta_-^\dagger \end{pmatrix}\mathcal{R}(x,\zeta),\label{eq:pregap06} \\
		K_{\text{bd}}(x,x)&=H(x)W(x)^\dagger=\sum_j h_je_j\hat{p}_j^\dagger\begin{pmatrix} I_d & s_j^{-1}\Delta_- \end{pmatrix}, \label{eq:pregap07}\\
		K_{\text{bd}}(x,x)^\dagger&=W(x)H(x)^\dagger=\sum_j \begin{pmatrix} I_d \\ s_j\Delta_-^\dagger \end{pmatrix}e_j\hat{p}_jh_j^\dagger. \label{eq:pregap08}
	\end{align}
	\indent Henceforth, we calculate
	\begin{align}
		F(x,s)F(x,s^*)^\dagger-\frac{mM_-+\mathrm{i}[\sigma_3,K(x,x)]}{ms},
	\end{align}
	which is part of the integrand of Eq.~(\ref{eq:gapxi001}).
	For brevity of description, in Eqs. (\ref{eq:pregap01}) and (\ref{eq:pregap02}), we write 
	$ F=[f_1+f_2+f_3]\mathrm{e}^{\mathrm{i}k(s)x},\ F^\dagger=[f_1'+f_2'+f_3']\mathrm{e}^{-\mathrm{i}k(s)x} $. Let us simplify $ FF^\dagger = \sum_{i,j=1}^3f_if_j' $. 
	First, we have
	\begin{align}
		f_1f_1'-\frac{M-}{s}=\begin{pmatrix}I_d & \\ & s^{-2}I_d \end{pmatrix},
	\end{align}
	which vanishes if we take a commutator with $ \sigma_3 $. (Even without a commutator, its integration vanishes if a principal value is taken.) Next, we consider the term $ f_2f_2' $. Using the partial fraction decomposition and Eqs. (\ref{eq:pregap03}) and (\ref{eq:pregap04}),
	\begin{align}
		f_2f_2'&=\frac{2\mathrm{i}}{m}\sum_j\left[ \frac{h_j}{s-s_j}\left( e_j\hat{p}_j^\dagger\frac{-2\mathrm{i}}{m}\sum_l\frac{e_l\hat{p}_lh_l^\dagger}{s_j-s_l^{-1}} \right) \right]-\frac{2\mathrm{i}}{m}\sum_l\left[ \left( \frac{-2\mathrm{i}}{m}\sum_j \frac{e_jh_j\hat{p}_j^\dagger}{s_j-s_l^{-1}}\right)\frac{h_l^\dagger}{s-s_l^{-1}} \right] \nonumber \\
		&=f_{\text{22a}}+f_{\text{22b}}+f_{\text{22c}}+f_{\text{22d}}+f_{\text{22e}}+f_{\text{22f}}
	\end{align}
	with
	\begin{align}
		f_{\text{22a}}=\frac{2\mathrm{i}}{m}\sum_j\frac{h_jh_j^\dagger}{s-s_j},\quad f_{\text{22b}}=\frac{2\mathrm{i}}{m}\sum_j\frac{h_je_j\hat{p}_j^\dagger}{s-s_j}\begin{pmatrix} I_d & s_j^{-1}\Delta_- \end{pmatrix},\quad f_{\text{22c}}=\frac{2\mathrm{i}}{m}\sum_j\frac{h_je_j\hat{p}_j^\dagger}{s-s_j}\int\frac{\mathrm{d}\zeta}{2\pi\mathrm{i}}\frac{\mathcal{R}(x,\zeta)}{s_j-\zeta}, \\
		f_{\text{22d}}=-\frac{2\mathrm{i}}{m}\sum_l\frac{h_lh_l^\dagger}{s-s_l^{-1}},\quad f_{\text{22e}}=-\frac{2\mathrm{i}}{m}\sum_l\begin{pmatrix} I_d \\ s_l \Delta_-^\dagger \end{pmatrix}\frac{e_l\hat{p}_lh_l^\dagger}{s-s_l^{-1}},\quad f_{\text{22f}}=-\frac{2\mathrm{i}}{m}\sum_l\int\frac{\mathrm{d}\zeta}{2\pi\mathrm{i}}\frac{s_l\tilde{R}(x,\zeta)}{s_l-\zeta}\frac{e_l\hat{p}_lh_l^\dagger}{s-s_l^{-1}}.
	\end{align}
	$f_{\text{22a}}$ and $f_{\text{22d}}$  provide a quantity relating to the filling rates of bound states:
	\begin{align}
		f_{\text{22a}}+f_{\text{22d}}=-\frac{4}{m}\sum_j\frac{h_jh_j^\dagger\sin\theta_j}{|s-s_j|^2},\quad \theta_j=\arg s_j.
	\end{align}
	Using Eqs.~(\ref{eq:pregap07}) and (\ref{eq:pregap08}),
	\begin{align}
		f_2f_1'+f_{\text{22b}}=\frac{-\mathrm{i}}{ms}K_{\text{bd}}(\sigma_3-1),\quad f_1f_2'+f_{\text{22e}}=\frac{\mathrm{i}}{ms}(\sigma_3-1)K_{\text{bd}}^\dagger.
	\end{align}
	Using Eqs.~(\ref{eq:pregap01}) and (\ref{eq:pregap06}),
	\begin{align}
		f_{\text{22c}}+f_2f_3'+f_1f_3'-\frac{\mathrm{i}}{ms}(\sigma_3-1)K_{\text{sc}}^\dagger=\int\frac{\mathrm{d}\zeta}{2\pi\mathrm{i}}\left[ F(x,\zeta)\mathrm{e}^{-\mathrm{i}k(\zeta)x}-\int\frac{\mathrm{d}\eta}{2\pi\mathrm{i}}\frac{\tilde{\mathcal{R}}(x,\eta)}{1-\zeta\eta} \right]\frac{\mathcal{R}(x,\zeta)}{s-\zeta}. \label{eq:pregap02plus00}
	\end{align}
	In the same way, from Eq.~(\ref{eq:pregap02}) with $ s=\zeta^{-1} $  and Eq. (\ref{eq:pregap05})
	\begin{align}
		f_{\text{22f}}+f_3f_2'+f_3f_1'+\frac{\mathrm{i}}{ms}K_{\text{sc}}(\sigma_3-1)=\int\frac{\mathrm{d}\zeta}{2\pi\mathrm{i}}\frac{\tilde{\mathcal{R}}(x,\zeta)}{1-s\zeta}\left[ F(x,\zeta^{-1*})^\dagger\mathrm{e}^{-\mathrm{i}k(\zeta)x}-\int\frac{\mathrm{d}\eta}{2\pi\mathrm{i}}\frac{\mathcal{R}(x,\eta)}{\zeta^{-1}-\eta} \right]. \label{eq:pregap02plus}
	\end{align}
	Finally,
	\begin{align}
		f_3f_3'=\iint\frac{\mathrm{d}\zeta}{2\pi\mathrm{i}}\frac{\mathrm{d}\eta}{2\pi\mathrm{i}}\frac{\tilde{\mathcal{R}}(x,\zeta)}{1-s\zeta}\frac{\mathcal{R}(x,\eta)}{s-\eta}. \label{eq:pregap02plus02}
	\end{align}
	Summarizing the above, let us write down $ FF^\dagger=\sum_{i,j=1}^3f_if_j $. The double integral terms in Eqs. (\ref{eq:pregap02plus00})-(\ref{eq:pregap02plus02}) cancel out. We also recall that $ K=K_{\text{sc}}+K_{\text{bd}}=K_{\text{sc}}^\dagger+K_{\text{bd}}^\dagger $. Thus,
	\begin{align}
		&FF^\dagger-\frac{m M_-+\mathrm{i}[\sigma_3,K]}{ms}\nonumber \\
		&=-\frac{4}{m}\sum_j\frac{h_jh_j^\dagger\sin\theta_j}{|s-s_j|^2}+\begin{pmatrix}I_d & \\ & s^{-2}I_d \end{pmatrix}+\int\frac{\mathrm{d}\zeta}{2\pi\mathrm{i}} F(x,\zeta)R(\zeta^*)^\dagger\Delta_-^\dagger\frac{F(x,\zeta^{-1*})^\dagger}{\zeta} \frac{(1-\zeta^2)}{(s-\zeta)(1-s\zeta)},
	\end{align}
	where $ \mathcal{R} $ and $ \tilde{\mathcal{R}} $ are eliminated by using their definition, Eq.~(\ref{eq:hilbertkernel}). 
	Let us integrate this expression by $ s $. Recalling that $ \zeta $ has an infinitesimal $ -\mathrm{i}0 $, we write $ \zeta=\zeta_R-\mathrm{i}0 $. The Sokhotski-Plemelj formula says
	\begin{align}
		\frac{1-\zeta^2}{(s-\zeta)(1-s\zeta)}=-\mathrm{i}\pi\left( \delta(s-\zeta_R)+\delta(s-\zeta_R^{-1}) \right)+\operatorname{\text{p.v.}}\left( \frac{1}{s-\zeta_R}-\frac{1}{s-\zeta_R^{-1}} \right), \label{eq:sokhotskiplemelj}
	\end{align}
	where p.v. denotes Cauchy's principal value. Thus,
	\begin{align}
		\left[\int_{-\infty}^0-\int_0^\infty  \right]\mathrm{d}s\frac{1-\zeta^2}{(s-\zeta)(1-s\zeta)} = 2\mathrm{i}\pi\operatorname{sgn}\zeta_R+2\log\zeta_R^2.
	\end{align}
	Using this, we obtain
	\begin{align}
		&\left[\int_{-\infty}^0-\int_0^\infty  \right]\frac{m\mathrm{d}s}{4\pi}\left( FF^\dagger-\frac{m M_-+\mathrm{i}[\sigma_3,K]}{ms} \right) \nonumber \\
		&= -\sum_j h_jh_j^\dagger\frac{2\theta_j-\pi}{\pi}+\int_{-\infty}^\infty\frac{m\mathrm{d}\zeta}{4\pi}\left(\operatorname{sgn}\zeta+\frac{\log \zeta^2}{\mathrm{i}\pi}\right)F(x,\zeta)R(\zeta)^\dagger\Delta_-^\dagger\frac{F(x,\zeta^{-1})^\dagger}{\zeta}.
	\end{align}
	In the last expression, $ -\mathrm{i}0 $ can be deleted (hence $ \zeta=\zeta^*=\zeta_R $) since the integral converges without this factor. 
%
	Thus, $ \Xi $ is finally written as
	\begin{align}
		&\Xi=\sum_j h_jh_j^\dagger\left(2\nu_j-\frac{2\theta_j}{\pi}\right)+\int_{-\infty}^\infty\frac{m\mathrm{d}s}{4\pi}\frac{\log s^2}{\mathrm{i}\pi}F(x,s)R(s)^\dagger\Delta_-^\dagger\frac{F(x,s^{-1})^\dagger}{s}, \label{eq:gapXifinal}
	\end{align}
	which is equivalent to Eqs. (\ref{eq:gapstep101}) and (\ref{eq:gapstep102}). We note that, in the previous work (Ref.~\cite{takahashinittaJLTP}), the delta function in Eq.~(\ref{eq:sokhotskiplemelj}) was not taken into account, and the final expression had an additional term, $ \operatorname{sgn}s $ or $ H(-s) $ (Eq. (21) or (23) of Ref.~\cite{takahashinittaJLTP}). If we correctly use Eq.~(\ref{eq:sokhotskiplemelj}), these terms vanish, as Eq.~(\ref{eq:gapXifinal}) shows. (This mistake does not change the main conclusion, namely, the reflectionless nature of the self-consistent solution.)
	
	\indent \textit{\underline{Step (II):}} We now prove the reflectionless nature of self-consistent potentials. First, we note that
	\begin{align}
		\tau \Xi_{\text{scattering}}^*\tau = \Xi_{\text{scattering}} 
	\end{align}
	for both $ \tau=\sigma_1  $ and $ \sigma_2 $ corresponding to symmetric and antisymmetric $ \Delta $'s. This is because $ \tau (Y_-\mathcal{P} Y_-^\dagger)^* \tau=Y_-\mathcal{P} Y_-^\dagger $ by Eqs. (\ref{eq:jostytau}), (\ref{eq:absym}), and (\ref{eq:abasym}). Thus the scattering part of the gap equation is the same for all three cases. As for bound states, following the result of Subsec.~\ref{subsec:boundsymantisym}, we can show
	\begin{alignat}{2}
		\sigma_1 H^*&=H \operatorname{diag}(s_1^{-1},\dots,s_n^{-1}), \quad&&(\text{for symmetric $ \Delta $}), \\
		\sigma_2 H^*&=H [(s_2^{-1}\sigma_y)\oplus(s_4^{-1}\sigma_y)\oplus\dotsb\oplus(s_n^{-1}\sigma_y)], \quad&&(\text{for antisymmetric $ \Delta $}),
	\end{alignat}
	where $ n $ is always even in the antisymmetric case, and pairs of bound states (see Subsec.~\ref{subsec:boundsymantisym}) are labeled as $ h_{2j-1} $ and $ h_{2j} $ (hence $ s_{2j-1}=s_{2j},\ \theta_{2j-1}=\theta_{2j} $), and $ \sigma_y $ is an ordinary $ 2\times 2 $ Pauli matrix. From this, we find
	\begin{alignat}{2}
		\sigma_1 \Xi_{\text{bound}}^*\sigma_1 &= \Xi_{\text{bound}} &&(\text{for symmetric $ \Delta $}), \\
		\sigma_2 \Xi_{\text{bound}}^*\sigma_2 &= 2H(\tilde{\mathcal{N}}-\Theta)H^\dagger &&(\text{for antisymmetric $ \Delta $}), \\
		\tilde{\mathcal{N}}&:=\operatorname{diag}(\nu_2,\nu_1,\nu_4,\nu_3,\dots,\nu_n,\nu_{n-1}).
	\end{alignat}
	$ \tilde{\mathcal{N}} $ is obtained by exchanging $ \nu_{2j-1} $ and $ \nu_{2j} $ in $ \mathcal{N} $. Thus, for non-symmetric, symmetric, and antisymmetric cases, the gap equation is
	\begin{align}
		[\sigma_3,\Xi_{\text{scattering}}+\tilde{\Xi}_{\text{bound}}]=0,
	\end{align}
	where $ \Xi_{\text{scattering}} $ is the same as above and  $ \tilde{\Xi}_{\text{bound}} $ is
	\begin{align}
		\tilde{\Xi}_{\text{bound}}=\begin{cases} \Xi_{\text{bound}} & \text{(for non-symmetric and symmetric $ \Delta $)}, \\ H(\mathcal{N}+\tilde{\mathcal{N}}-2\Theta)H^\dagger & \text{(for antisymmetric $\Delta$)}. \end{cases}
	\end{align}
	Let us consider the integral
	\begin{align}
		\int_{-\infty}^\infty\mathrm{d}x Y_+(x,s_1)^\dagger [\sigma_3,\Xi_{\text{scattering}}+\tilde{\Xi}_{\text{bound}}] Y_+(x,s_1)=0,\quad s_1\in\mathbb{R}.
	\end{align}
	From Eq.~(\ref{eq:constwjnew02}), it is trivial that
	\begin{align}
		\int_{-\infty}^\infty\mathrm{d}x Y_+(x,s_1)^\dagger [\sigma_3,\tilde{\Xi}_{\text{bound}}] Y_+(x,s_1)=0.
	\end{align}
	As for the scattering parts, we use the formula (\ref{eq:ortho007}) proved in Appendix~\ref{app:fourortho}. For convenience of comparison, we rewrite the integration variable as $ s\rightarrow s_2 $ in Eq.~(\ref{eq:gapstep102}). (Note that the $ s_1,s_2 \in \mathbb{R} $ considered here is different from those of bound states on $ |s|=1 $.) In Eq.~(\ref{eq:ortho007}), let us set
	\begin{align}
		\mathcal{O}=\mathcal{P}(s_2), \label{eq:ortho00701}
	\end{align}
	then it satisfies
	\begin{align}
		&\{\sigma_3,\mathcal{P}(s_2)\}=\{\sigma_3,S(s_2)^\dagger\mathcal{P}(s_2)S(s_2)\}=0,  \label{eq:ortho00702} \\
		&M_-\mathcal{P}(s_2)M_-=\mathcal{P}(s_2^{-1}). \label{eq:ortho00703}
	\end{align}
	By Eq.~(\ref{eq:ortho00702}), the p.v. part in Eq. (\ref{eq:ortho007}) vanishes. By Eqs.~(\ref{eq:scatMpMm}) and (\ref{eq:ortho00703}), the first and last terms in Eq.~(\ref{eq:ortho007}) are shown to provide the same terms after $s_2$-integration. After a calculation, we obtain
	\begin{align}
		&\int_{-\infty}^\infty\mathrm{d}x Y_+(x,s_1)^\dagger [\sigma_3,\Xi_{\text{scattering}}] Y_+(x,s_1)=0\\
		\leftrightarrow\quad&[\sigma_3,S(s_1)^\dagger\mathcal{P}(s_1)S(s_1)]+S(s_1)^\dagger[\sigma_3,\mathcal{P}(s_1)]S(s_1)=0,
	\end{align}
	which gives
	\begin{align}
		B(s_1)=0 \quad \leftrightarrow\quad R(s_1)=0 \qquad \text{for $ s_1\in \mathbb{R} $.}  \label{eq:thegapfin001}
	\end{align} 
	Thus, the self-consistent potential is reflectionless. \\ 
	\indent \textit{\underline{Step (III):}} The final step, the solution of $ [\sigma_3,\tilde{\Xi}_{\text{bound}}]=0 $, is trivial. $ [\sigma_3,h_jh_j^\dagger] $ for $ j=1,\dots, n $ are linearly independent, since  they have different decay rates for different eigenvalues and the coefficient vectors are chosen to be orthogonal for degenerate eigenvalues. Thus, we conclude that
	\begin{alignat}{2}
		\mathcal{N}&=\Theta \quad &&\text{for non-symmetric and symmetric cases}, \\
		\mathcal{N}+\tilde{\mathcal{N}}&=2\Theta  \quad &&\text{for the antisymmetric case}, \label{eq:thegapfin003}
	\end{alignat}
	which is consistent with Ref.~\cite{arxiv1509.04242}. Thus, the solution of the gap equation is given by Eqs.~(\ref{eq:thegapfin001})-(\ref{eq:thegapfin003}), which proves the reflectionless nature of self-consistent potentials.

\section{Asymptotics of $n$-soliton solutions}\label{sec:asymptotics}
	In this section, we consider reflectionless potentials, i.e.,  $ n $-soliton solutions, in detail. We derive asymptotic formulae when solitons are sufficiently separated from each other. 
	Although we do not need concrete expressions and asymptotic formulae for multi-soliton solutions when we solve the gap equation in Sec.~\ref{sec:gap}, the formulae shown in this section are helpful to grasp the physical picture of multi-soliton states.

\subsection{Reflectionless solution}\label{subsec:asymptotic}
	\indent Let us solve the GLM equation for the reflectionless case $ R(s)=0 $ for real $ s $. 
	We assume that there are  $ n $ bound states, and their eigenvalues are given by $ \epsilon(s_1),\dots, \epsilon(s_n),\ s_i\in\mathbb{H} $. For convenience, we write
	\begin{align}
		\kappa_j=-\mathrm{i}k(s_j),\ c_j=\sqrt{\kappa_j}\mathrm{e}^{-\kappa_kx_j},\ e_j(x)=c_j\mathrm{e}^{-\mathrm{i}k(s_j)x}=\sqrt{\kappa_j}\mathrm{e}^{\kappa_j(x-x_j)}.
	\end{align}
	As we will see below, the position of the $ j $-th soliton $ X_j $ is expressed as $ X_j=x_j+ $~(additive constant arising from interaction between solitons).   The kernel  $ K $ and bound states $ H $ are 
	\begin{align}
		K(x,y)&=H(x)W(y)^\dagger,\\
		H(x)&=(h_1(x),\dots,h_n(x)),\\
		W(x)&=\begin{pmatrix} e_1(x)\hat{p}_1 & \dots & e_n(x)\hat{p}_n \\ s_1e_1(x)\Delta_-^\dagger\hat{p}_1 &\dots& s_ne_n(x)\Delta_-^\dagger\hat{p}_n  \end{pmatrix}.
	\end{align}
	Defining the Gram matrix $ G(x)=\int_{-\infty}^x\mathrm{d}yW(y)^\dagger W(y) $, whose matrix components are given by Eq.~(\ref{eq:thegram}), the GLM equation reduces to 
	\begin{align}
		H(x)+W(x)+H(x)G(x)=0. \label{eq:solasymfn01}
	\end{align}
	Thus, the bound states are given by $ H=-W(I+G)^{-1} $. 
	The scattering states are given as follows. The integral representation of the left Jost function is given by Eq.~(\ref{eq:JostkernelKl0}) or (\ref{eq:JostkernelKr0}). For convenience of comparison with Ref.~\cite{arxiv1509.04242}, we define $ F(x,s)=s^{-1}F_-(x,s)\Delta_-^\dagger $. Then the scattering state is given by
	\begin{align}
		F(x,s)=\left[ \begin{pmatrix} I_d \\ s^{-1}\Delta_-^\dagger \end{pmatrix}+\frac{2\mathrm{i}}{m}\sum_{j=1}^n\frac{h_j(x)e_j(x)\hat{p}_j^\dagger}{s_j-s} \right]\mathrm{e}^{\mathrm{i}k(s)x},\quad s\in\mathbb{R}.
	\end{align}
	 $ F(x,s)\mathrm{e}^{-\mathrm{i}\epsilon(s)t} $ corresponds to $ F(x,t,s) $ of Ref.~\cite{arxiv1509.04242}. The gap function $ \Delta $ is given by Eq.~(\ref{eq:kernelKdiff215}), and, in the current case,
	 \begin{align}
	 	\Delta(x)=m\Delta_--2\mathrm{i}(e_1(x)\hat{p}_1,\dots,e_n(x)\hat{p}_n)(I_n+G(x))^{-1}\begin{pmatrix} s_1^{-1}e_1(x)\hat{p}_1^\dagger \\ \vdots \\ s_n^{-1}e_n(x)\hat{p}_n^\dagger \end{pmatrix}\Delta_-.
	 \end{align}

\subsection{One-soliton solution}
	For later convenience, we first introduce the notations for the most basic one-soliton solution in the one-component system. We write $ s_i=\mathrm{e}^{\mathrm{i}\theta_i} $, and hence $ \kappa_i=m\sin\theta_i $. Then, let us define
	\begin{align}
		f_{\text{basic}}(x-x_1,s_1)&:=-\frac{e_1}{1+\frac{e_1^2}{m\sin\theta_1}}= -\frac{\sqrt{\kappa_1}}{2}\frac{1}{\cosh\kappa_1(x-x_1)}, \\ 
		\Delta_{\text{basic}}(x-x_1,s_1)&:=m\frac{1+\frac{e_1^2s_1^{-2}}{m\sin\theta_1}}{1+\frac{e_1^2}{m\sin\theta_1}}=m\mathrm{e}^{-\mathrm{i}\theta_1}\left[ \cos\theta_1-\mathrm{i}\sin\theta_1\tanh \kappa_1 (x-x_1) \right]. \label{eq:onesoldenot}
	\end{align}
	Using the above notations, if $ \Delta(x) $ is $ d\times d $, the one-soliton solution located at $ x=x_1 $ is as follows. The bound state is given by 
	\begin{align}
		h_1(x)=f_{\text{basic}}(x-x_1,s_1)\begin{pmatrix} \hat{p}_1 \\ s_1\Delta_-^\dagger\hat{p}_1 \end{pmatrix},
	\end{align}
	and $ \Delta(x) $ is
	\begin{align}
		\Delta(x) = \left[m(I_d-\hat{p}_1\hat{p}_1^\dagger)+\Delta_{\text{basic}}(x-x_1,s_1)\hat{p}_1\hat{p}_1^\dagger\right]\Delta_-. \label{eq:onesolmatrix}
	\end{align}
	If we introduce a unitary matrix $ D_1 $ such that $ \hat{p}_1=D_1(1,0,\dots,0)^T $, 
	\begin{align}
		\Delta(x) = D_1 \operatorname{diag}(\Delta_{\text{basic}}(x-x_1,s_1),m,\dots,m)D_1^\dagger\Delta_-.
	\end{align}
	Thus, the structure of the one-soliton solution is actually very simple;  that is, if we choose a basis such that $ \Delta $ becomes diagonal, one of diagonal elements is given by the one-soliton solution of the one-component system, and others are simply given by a constant. However, if the number of solitons increases, due to the arbitrariness of the choice of $ \hat{p}_1,\hat{p}_2,\dots \hat{p}_n $, the solitons have various ``angles'', and hence the treatment becomes less easy compared to the one-component problem. However, if the solitons are well separated, each soliton can be approximated by the form of the one-soliton expression (\ref{eq:onesolmatrix}). In the next subsection, we derive such formulae.
\subsection{Approximate formulae for isolated solitons}
	\begin{figure}[tb]
		\begin{center}
		\includegraphics[scale=.55]{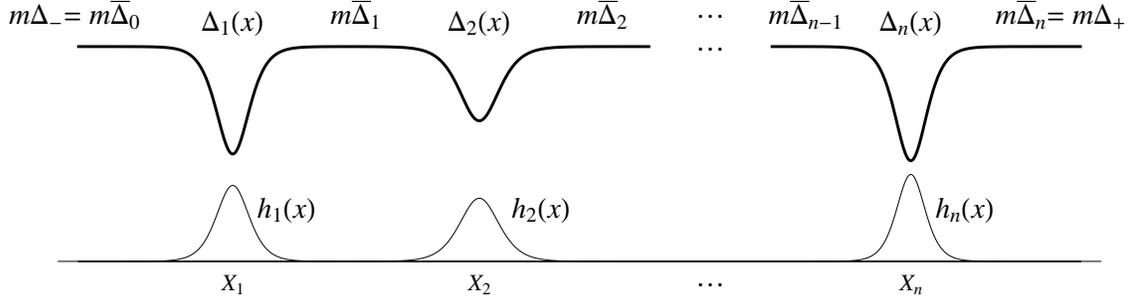}
		\caption{\label{fig:solasyms} Schematic picture to explain the meaning of $ \Delta_j(x) $'s, $ \bar{\Delta}_j $'s, and $ h_j(x) $'s.  If solitons are sufficiently separated from other solitons,  $ \Delta(x) $ near each soliton is approximated by the one-soliton expression [Eq.~(\ref{eq:onesolmatrix})]. $ \Delta_j(x) $ is the expression near the $ j $-th soliton and $ X_j $ represents its position. $ h_j(x) $ represents the corresponding bound state. $ m\bar{\Delta}_j $ represents the constant value in the uniform region between $ j $-th and $ (j+1) $-th soliton. These quantities are determined by $ s_i,\ \hat{p}_i, $ and $  x_i $'s [Eqs.~(\ref{eq:solasymfun22})-(\ref{eq:bardeltarec41})].}
		\end{center}
	\end{figure}
	In this subsection, we provide approximate expressions of multi-soliton solutions when solitons are well isolated. The aim is the determination of $ \bar{\Delta}_j $'s and $ \Delta_j(x) $'s in Fig.~\ref{fig:solasyms}.   \\
	\indent Now let us assume that solitons are sufficiently separated from each other, and assume $ x_1 \ll x_2 \ll \dotsb \ll x_n $. We are interested in an approximate expression of $ j $-th soliton near $ x \simeq x_j $. Let us consider how Eq.~(\ref{eq:solasymfn01}) is simplified if we focus only on the vicinity of $ j $-th soliton $ x\simeq x_j $. The bound state $ h_k $ rapidly decreases far from the $ k $-th soliton; thus we can set $ H \simeq (0,\dots,h_j,0,\dots,0) $. As for $ W $, while $ e_1,\dots,e_j $ should be taken into account since they grow exponentially, the remaining $ e_{k>j}(x) $ are ignorable. In the last term $ HG $, though $ h_{k<j}(x) $ are exponentially small, the product of $ h_k $ and $ e_k $ gives an $O(1)$ term; thus they should be kept. Summarizing, Eq.~(\ref{eq:solasymfn01}) is approximated by
	\begin{align}
		H_j(\hat{\xi}_j\hat{\xi}_j^\dagger+G_j)+W_j=0 \label{eq:solasymfun1}
	\end{align}
	where $ H_j $ and $ W_j $ are  $ (2d)\times j $ matrices made by the first $ j $ columns of $ H $ and $ W $, and $ G_j $ is a top-left $ j\times j $ submatrix of $ G $, and  $ \hat{\xi}_j=(0,\dots,0,1)^T $ is a  $ j $-component unit column vector. 
	The kernel  $ K $ in this approximation is given by $ K(x,y)=H_j(x)W_j(y)^\dagger $ and  $ \Delta $ is given by $ \Delta(x)=m\Delta_-+2\mathrm{i}K_{12}(x,x) $ with $ K_{12} $ being a top-right $ d\times d $ block of $ K $. \\
	\indent For brevity, we introduce a few more notations.  $ \mathcal{S}_j:=\operatorname{diag}(s_1,\dots,s_j) $, $ \mathcal{E}_j(x):=\operatorname{diag}(e_1(x),\dots,e_j(x)) $, and 
	\begin{align}
		Q_j:=m \mathcal{E}_j^{-1}G_j\mathcal{E}_j^{-1},\quad [Q_j]_{kl}=\frac{2\mathrm{i}\hat{p}_k^\dagger\hat{p}_l}{s_k-s_l^{-1}}. \label{eq:solasym37}
	\end{align}
	$ Q_j $ is positive definite since $ G_j $ is so. Let $ \tilde{Q}_j $ be a  $ (j-1)\times j $ matrix such that the  $ j $-th row in $ Q_j $ is deleted. \\
	\indent Then, solving Eq.~(\ref{eq:solasymfun1}), we obtain the following formulae: The approximate expression for the $ j $-th bound state $ h_j(x) $ and the $j$-th soliton $ \Delta_j(x) $ are given by 
	\begin{align}
		h_j(x)&=f_{\text{basic}}(x-X_j,s_j)\begin{pmatrix} \hat{q}_j \\ \hat{r}_j \end{pmatrix}, \label{eq:solasymfun22} \\
		\Delta_j(x)&=\bar{\Delta}_{j-1}\left( m(I_d-\hat{r}_j\hat{r}_j^\dagger)+\Delta_{\text{basic}}(x-X_j,s_j)\hat{r}_j\hat{r}_j^\dagger \right)=\left( m(I_d-\hat{q}_j\hat{q}_j^\dagger)+\Delta_{\text{basic}}(x-X_j,s_j)\hat{q}_j\hat{q}_j^\dagger \right)\bar{\Delta}_{j-1},
	\end{align}
	where $ X_j:=x_j+y_j $ and the position shift $ y_j $, the coefficient vectors $ \hat{q}_j,\hat{r}_j $, and unitary matrices $ \bar{\Delta}_j $ are given by
	\begin{align}
		&\mathrm{e}^{-2\kappa_j y_j}:=\sin\theta_j\frac{\det Q_j}{\det Q_{j-1}}, \quad (j\ge 2), \\
		&\hat{q}_j:=\frac{\det\begin{pmatrix} \tilde{Q}_j \\ \hat{p}_1 \ \cdots \ \hat{p}_j \end{pmatrix}}{\sqrt{\sin\theta_j \det Q_{j-1}\det Q_j}},\quad \hat{r}_j:=\frac{\det\begin{pmatrix} \tilde{Q}_j \\ s_1\Delta_-^\dagger \hat{p}_1 \ \cdots \ s_j\Delta_-^\dagger \hat{p}_j \end{pmatrix}}{\sqrt{\sin\theta_j \det Q_{j-1}\det Q_j}}, \quad (j\ge 2), \\
		&\bar{\Delta}_j:=\Delta_--2\mathrm{i}(\hat{p}_1,\dots,\hat{p}_j)Q_j^{-1}\begin{pmatrix} s_1^{-1}\hat{p}_1^\dagger \\ \vdots \\ s_j^{-1}\hat{p}_j^\dagger \end{pmatrix}\Delta_-=\Delta_--2\mathrm{i}\sum_{k=1}^j(\sin\theta_k)\hat{q}_k\hat{r}_k^\dagger, \quad (j\ge 1).
	\end{align}
	Note that $ \hat{q}_j,\hat{r}_j $ are normalized: $\hat{q}_j^\dagger\hat{q}_j=\hat{r}_j^\dagger\hat{r}_j=1$. We define $ y_1=0 $, $ \hat{q}_1=\hat{p}_1 $, $ \hat{r}_1=s_1\Delta_-^\dagger\hat{p}_1 $, and $ \bar{\Delta}_0=\Delta_- $. It also follows that $ \bar{\Delta}_n=\Delta_+ $. The position shift $ y_j $, which arises from soliton scattering, is always non-negative $ y_j\ge 0 $, since  $ \mathrm{e}^{-2\kappa_j y_j}\le 1 $. (Because of the fact that $ Q_j $'s are positive definite and $ \frac{1}{\sin\theta_j}=[Q_j]_{jj} $ and the general inequality $ \det H \le \det H_{11}\det H_{22} $ where $ H $ is positive-definite hermitian and $ H_{ii} $ are its square diagonal blocks.) 
	$ \hat{q}_j,\ \hat{r}_j,\ \bar{\Delta}_{j-1}, $ and $ \bar{\Delta}_j $  are related by
	\begin{align}
		&\bar{\Delta}_j\hat{r}_j=s_j^{-1}\hat{q}_j,\quad \bar{\Delta}_{j-1}\hat{r}_j=s_j\hat{q}_j, \\
		&\bar{\Delta}_{j-1}=\bar{\Delta}_j+(s_j-s_j^{-1})\hat{q}_j\hat{r}_j^\dagger. \label{eq:bardeltarec01}
	\end{align}
	$ m\bar{\Delta}_j $ represents the constant value of the potential between the $ j $-th and $ (j+1) $-th soliton, $ \Delta_j(x) $ and $ \Delta_{j+1}(x) $. 
%
	We can obtain a few more recurrence relations for $ \bar{\Delta}_j $'s other than Eq.~(\ref{eq:bardeltarec01}). If we define
	\begin{align}
		U_j=I_d+(s_j^{-1}-1)\hat{q}_j\hat{q}_j^\dagger=\mathrm{e}^{-\mathrm{i}\theta_j \hat{q}_j\hat{q}_j^\dagger},\quad V_j=I_d+(s_j^{-1}-1)\hat{r}_j\hat{r}_j^\dagger=\mathrm{e}^{-\mathrm{i}\theta_j \hat{r}_j\hat{r}_j^\dagger},
	\end{align} 
	then Eq. (\ref{eq:bardeltarec01}) is equivalent to
	\begin{align}
		\bar{\Delta}_j=U_j^2\bar{\Delta}_{j-1}=\bar{\Delta}_{j-1}V_j^2=U_j\bar{\Delta}_{j-1}V_j.
	\end{align}
	Repeatedly using this, we obtain (recall that $ \bar{\Delta}_0=\Delta_- $.)
	\begin{align}
		\bar{\Delta}_j=U_j^2\dotsm U_1^2\bar{\Delta}_0=\bar{\Delta}_0V_1^2\dotsm V_j^2=U_j\dotsm U_1\bar{\Delta}_0V_1\dotsm V_j. \label{eq:bardeltarec41}
	\end{align}
	\indent Here, we sketch how to derive the above formulae (\ref{eq:solasymfun22})-(\ref{eq:bardeltarec41}). We only show three key linear-algebraic formulae and omit a lengthy calculation. First,
	\begin{align}
		1+y^\dagger A^{-1} x &= \frac{\det (A+xy^\dagger)}{\det A} \label{detlemma1},
	\end{align}
	which is one of the Sherman-Morrison-Woodbury type formulas. 
	Next, 
	\begin{align}
		\hat{p}_k^\dagger\hat{p}_l=\frac{s_k-s_l^{-1}}{2\mathrm{i}}[Q_j]_{kl}=\frac{[\mathcal{S}_jQ_j-Q_j\mathcal{S}_j^{-1}]_{kl}}{2\mathrm{i}}, \label{eq:pigramtoQ}
	\end{align}
	which is a slight rewriting of Eq.(\ref{eq:solasym37}), but very convenient. Finally, let $ A $ be an invertible $ n\times n $ matrix, and  $ B $ an  $ (n-1)\times (n-1) $ matrix such that the $ n $-th column and row of $ A $ are deleted. Let $ C $  be a cofactor matrix of $ A $ satisfying $ AC=CA=(\det A)I_n $. 
	(Note: the ``cofactor matrix'' may represent $ C^T $ in other references.) 
	From Jacobi's formula\footnote{Also called the Jacobi-Desnanot formula, the Lewis Carroll identity, etc.} \cite{Satake},
	\begin{align}
		[B^{-1}]_{ij}=\frac{C_{nn}C_{ij}-C_{in}C_{nj}}{\det A\det B},\ \quad 1\le i,j \le n-1. \label{lewiscarroll}
	\end{align}
	We note that $ C_{nn}=\det B $ by definition. Using this formula, we can prove
	\begin{align}
		A^{-1}-\begin{pmatrix} B^{-1} & 0 \\ 0 & 0 \end{pmatrix}=\frac{1}{\det A \det B}\begin{pmatrix} C_{1n} \\ C_{2n} \\ \vdots \\ C_{nn} \end{pmatrix}\begin{pmatrix} C_{n1} & C_{n2} & \cdots & C_{nn} \end{pmatrix}. \label{lewiscarroll2}
	\end{align}
	The examples of usage of Eqs. (\ref{detlemma1})-(\ref{lewiscarroll2}) are as follows. For example, when we solve Eq.~(\ref{eq:solasymfun1}) by Cramer's rule, we need $ \det (G_j+\hat{\xi}_j\hat{\xi}_j^\dagger) $. This is rewritten as $ \det (G_j+\hat{\xi}_j\hat{\xi}_j^\dagger)=(1+\hat{\xi}_j^\dagger G_j^{-1}\hat{\xi}_j)\det G_j=\det G_j+\det G_{j-1} \propto \frac{e_j^2}{m}\det Q_j+\det Q_{j-1} $, which provides the denominator of $ f_{\text{basic}} $ and $ \Delta_{\text{basic}} $. 
	As another example, the fact that $ \hat{q}_j $ is normalized is shown as follows. Let $ C $ denote a cofactor matrix of $ Q_j $. Then, $ \hat{q}_j^\dagger\hat{q}_j=\frac{1}{\sin\theta_j \det Q_{j-1}\det Q_j} \sum_{k,l}C_{jk}C_{lj}\hat{p}_k^\dagger \hat{p}_l=\frac{1}{\sin\theta_j}\sum_{k,l}\left( [Q_j^{-1}]_{lk}-\Bigl[ \Bigl( \begin{smallmatrix} Q_{j-1}^{-1} & 0 \\ 0&0 \end{smallmatrix} \Bigr) \Bigr]_{lk} \right)\frac{[S_jQ_j-Q_jS_j^{-1}]_{kl}}{2\mathrm{i}}=\frac{1}{2\mathrm{i}\sin\theta_j}\Bigl( \operatorname{tr}(Q_j^{-1}\mathcal{S}_jQ_j-Q_j^{-1}Q_j\mathcal{S}_j^{-1})-\operatorname{tr}(Q_{j-1}^{-1}\mathcal{S}_{j-1}Q_{j-1}-Q_{j-1}^{-1}Q_{j-1}\mathcal{S}_{j-1}^{-1}) \Bigr)=1 $. Other relations are also shown in a similar manner.

\section{Stationary-class potentials and the matrix NLS equation}\label{sec:matrixnls}
	\indent In Ref.~\cite{arxiv1509.04242}, it is said that the stationary-class potentials are the same as the ``snapshots'' for every time $ t $ of the soliton solutions for the matrix NLS equation. Here we demonstrate it. \\
	\indent Instead of the bare equation $ \mathrm{i}\Delta_t=-\Delta_{xx}+2\Delta\Delta^\dagger\Delta $, by gauge transformation $ \Delta \rightarrow \Delta\mathrm{e}^{-2\mathrm{i}m^2t},\ m>0 $, we discuss
	\begin{align}
		\mathrm{i}\Delta_t=-2m^2\Delta-\Delta_{xx}+2\Delta\Delta^\dagger\Delta, \label{eq:matrixNLS001}
	\end{align}
	then it has the finite-density uniform solution $ \Delta(t,x)=m\Delta_- $, where $ \Delta_- $ is a unitary matrix. 
	The Lax pair for Eq.~(\ref{eq:matrixNLS001}) is
	\begin{align}
		\mathcal{L}&=\begin{pmatrix} -\mathrm{i}\partial_x & \Delta \\ \Delta^\dagger & \mathrm{i}\partial_x \end{pmatrix}, \\
		\mathcal{B}&=\begin{pmatrix} 2\partial_x^2+(m^2-\Delta\Delta^\dagger)  & 2\mathrm{i}\Delta\partial_x+\mathrm{i}\Delta_x \\ 2\mathrm{i}\Delta^\dagger\partial_x+\mathrm{i}\Delta_x^\dagger &-2\partial_x^2-(m^2-\Delta^\dagger\Delta) \end{pmatrix},
	\end{align}
	and the matrix NLS equation is equivalent to the Lax equation $ \mathrm{i}\mathcal{L}_t=[\mathcal{L},\mathcal{B}] $.  The Lax equation reduces to the linear problem
	\begin{align}
		\mathcal{L}w&=\epsilon w, \label{eq:laxlin01}\\
		\mathrm{i}w_t&=-\mathcal{B}w  \label{eq:laxlin02}
	\end{align}
	with isospectral condition $ \epsilon_t=0 $. Rewriting this to the equivalent AKNS system is straightforward.\\
	\indent Let us find the time-dependence of the scattering matrix $ A(s) $ and $ B(s) $ of the Jost function. Let us assume that $ \Delta(t,x) $ obeys the matrix NLS equation and satisfies the boundary condition $ \Delta(t,\pm\infty)=m\Delta_\pm $. We write $ \Delta(0,x)=\Delta(x) $.  Let $ \tilde{Y}_+(t,x,s) $ be a solution of Eqs. (\ref{eq:laxlin01}) and (\ref{eq:laxlin02}), with $ \epsilon=\epsilon(s) $ and the initial condition $ w(t=0,x)=Y_+(x,s) $. Though $ \mathcal{B} $ is time-dependent near the origin, its asymptotic form at spatial infinities
	\begin{align}
		\mathcal{B}_\pm :=\lim_{x\rightarrow\pm\infty}\mathcal{B}=\begin{pmatrix} 2\partial_x^2 & 2\mathrm{i}m\Delta_\pm \partial_x \\ 2\mathrm{i}m\Delta_\pm^\dagger\partial_x & -2\partial_x^2 \end{pmatrix}
	\end{align}
	does not depend on time. So the asymptotic form of $ \tilde{Y}_+ $ can be easily obtained. We can soon check
	\begin{align}
		-\mathcal{B}_\pm\Psi_\pm(x,s)=2\epsilon(s)k(s)\Psi_\pm(x,s)\sigma_3,
	\end{align}
	so therefore the solution of Eq. (\ref{eq:laxlin02}) at $ x=\pm\infty $ is given by
	\begin{align}
		\tilde{Y}_+(t,x,s)\rightarrow\begin{cases} \Psi_-(x,s)\mathrm{e}^{-2\mathrm{i}\epsilon(s)k(s)\sigma_3 t}S(s) & (x\rightarrow-\infty), \\ \Psi_+(x,s)\mathrm{e}^{-2\mathrm{i}\epsilon(s)k(s)\sigma_3 t} & (x\rightarrow+\infty). \end{cases}
	\end{align}
	Now let us define the time-dependent Jost function $ Y_+(t,x,s) $  by the asymptotic form:
	\begin{align}
		Y_+(t,x,s)\rightarrow \Psi_+(x,s), \quad (x\rightarrow+\infty),
	\end{align}
	and define the time-dependent scattering matrix $ S(t,s) $  by 
	\begin{align}
		Y_+(t,x,s)\rightarrow \Psi_-(x,s)S(t,s), \quad (x\rightarrow-\infty).
	\end{align}
	 $ A(t,s) $ and $ B(t,s) $ are defined by the top-left and bottom-left blocks of $ S(t,s) $, as with Eq.~(\ref{eq:SsAsBs}). Comparing the asymptotic forms for $ x\rightarrow+\infty $, we obtain the relation $ \tilde{Y}_+(t,x,s)=Y_+(t,x,s)\mathrm{e}^{-2\mathrm{i}\epsilon(s)k(s)\sigma_3 t} $. Furthermore, comparing the asymptotic forms at $ x\rightarrow-\infty $, we obtain the time-evolution of the scattering matrix
	\begin{align}
		S(t,s)=\mathrm{e}^{-2\mathrm{i}\epsilon(s)k(s)\sigma_3 t} S(s) \mathrm{e}^{2\mathrm{i}\epsilon(s)k(s)\sigma_3 t},
	\end{align} 
	which means
	\begin{align}
		A(t,s)=A(s),\quad B(t,s)=\mathrm{e}^{4\mathrm{i}\epsilon(s)k(s)t}B(s).
	\end{align}
	The coefficient $ c_j $ of the bound states is related by $ P_jP_j^\dagger =  \sum_i c_{ji}^2\hat{p}_i\hat{p}_i^\dagger \propto B(s_j)C_j $ (Subsec.~\ref{subsubsec:Hj}, Eq.(\ref{eq:PPdaggerBC})). Thus, its time dependence is given by
	\begin{gather}
		c_j(t)=\mathrm{e}^{-2\epsilon_j\kappa_jt}c_j,\label{eq:matrixnslcjt} \\
		 \epsilon_j=\epsilon(s_j)=m\cos\theta_j,\ \kappa_j=-\mathrm{i}k(s_j)=m\sin\theta_j.
	\end{gather}
	For these $ A(t,s),\ B(t,s), $ and $ c_j(t) $, solving the inverse problem, i.e., the GLM equation (Subsec.~\ref{subsec:glm}) for each time $ t $, we obtain the time evolution of $ \Delta(t,x) $. This is a traditional result of soliton theory based on the IST. \\ 
	\indent Because of Eq.~(\ref{eq:matrixnslcjt}), the multi-soliton solution of the matrix NLS equation can be obtained by the replacement
	\begin{align}
		e_j(x)=\sqrt{\kappa_j}\mathrm{e}^{\kappa_j(x-x_j)} \quad\rightarrow\quad e_j(t,x)=\sqrt{\kappa_j}\mathrm{e}^{\kappa_j(x-x_j-2\epsilon_j t)}.
	\end{align}
	in the equations of Subsec.~\ref{subsec:asymptotic}. $ h_j $'s  also have a time-dependence via $ e_j(t,x) $'s, since they are the solution of Eq.~(\ref{eq:solasymfn01}).   We must emphasize that such a time-evolution rule is different from the time-dependent BdG problem \cite{arxiv1509.04242}. The time evolution of the matrix NLS equation can be simply obtained by ``sliding'' the position parameter $ x_j $ by velocity $ V_j=2\epsilon_j $ in the stationary-class potentials. These facts indicate that investigation of time-dependent BdG solutions is not a mere revisit of the well-known solutions of soliton equations.

	Using the result of Sec.~\ref{sec:asymptotics}, we will be able to obtain an asymptotic form of the multi-soliton solution for $ t\rightarrow\pm\infty $. For $ t=+\infty $, the order of solitons is just equal to the order of their velocities. For $ t=-\infty $, the order is reversed. For these soliton orderings, the formulae of Sec.~\ref{sec:asymptotics} can be applied. Here, however, we do not go into any more detail on this topic.

\section{Summary and perspective} \label{sec:summary}
	In this paper, we have solved the gap equation for the matrix BdG systems. Sections \ref{sec:ist} and \ref{sec:gap} include the main result of this paper. Namely, we have determined the self-consistent condition [Eqs. (\ref{eq:thegapfin001})-(\ref{eq:thegapfin003})], in which the reflection coefficients of the scattering states must vanish, and the bound states must have partial filling rates. The filling rates are consistent with Ref.~\cite{arxiv1509.04242}. Sections \ref{sec:asymptotics} and \ref{sec:matrixnls} treat supplementary but important topics. In Sec.~\ref{sec:asymptotics}, we have provided approximate expressions for multi-soliton solutions that are valid when solitons are well separated.
 In Sec.~\ref{sec:matrixnls}, solving the initial-value problem of the matrix NLS equation, we have shown that the stationary-class potentials are snapshots of the soliton solutions of the matrix NLS equation, as first mentioned in Ref.~\cite{arxiv1509.04242}. \\
	\indent The mathematical expressions and techniques constructed in this paper will play an important role in guessing the time-dependent and finite-reflection solutions, which is the largest set depicted in Fig.~\ref{fig:intro}. The soliton solutions constructed in this paper are also useful for finding the solutions in the presence of piecewise constant potentials, and hence, the investigation of the physics of multicomponent superconductors in inhomogeneous systems such as edge states and Josephson effects (e.g., using the formula in Ref. \cite{FURUSAKI1991299}) will also be an important future work.

%

\appendix

\section{Existence of integral representation of Jost functions}\label{app:Kintegral}

	The most important key to the GLM formalism is introducing the integral representation of the Jost function [Eq.~(\ref{eq:JostkernelKl0}) or (\ref{eq:JostkernelK})], but its justification is often omitted.\footnote{It should be emphasized that Eq.~(\ref{eq:JostkernelK}) is \textit{not} a mere rewriting of the differential equation as an equivalent integral equation. If we rewrite the differential equation (\ref{eq:mZS}) as an integral equation with boundary condition $ Y_-(-\infty,s)=\Psi_-(-\infty,s) $, the resultant equation is 
	\begin{align*}
		Y_-(x,s)=\Psi_-(x,s)+\Psi_-(x,s)\int_{-\infty}^x\mathrm{d}y\Psi_-(y,s)^{-1}(U(y)-M)Y_-(y,s). 
	\end{align*}
	The kernel of this equation, however, includes $ s $-dependence (i.e., $\epsilon$-dependence). On the other hand, the kernel $ K(x,y) $ in Eq. (\ref{eq:JostkernelK}) has no $s$-dependence.
	Thus the existence of such a representation is not trivial. 
	}
	 To justify this, we must show the absolute convergence of the iterative solution of the Volterra integral equation which $ K(x,y) $ satisfies. 
	 In Sec. I-8 of Ref.~\cite{FaddeevTakhtajan}, it is described by a text. 
	 Here we supplement this statement for matrix-generalized ZS operator [Eq.~(\ref{eq:mZS})]. \\ 
	\indent Let $ K_1(x,y):=\frac{1}{2}(K(x,y)+\sigma_3 K(x,y)\sigma_3),\ K_2(x,y):=\frac{1}{2}(K(x,y)-\sigma_3 K(x,y)\sigma_3) $,
	and $U(x)=\left(\begin{smallmatrix} & -\mathrm{i}\Delta(x) \\ \mathrm{i}\Delta(x)^\dagger & \end{smallmatrix}\right),\ M=U(-\infty)=-\mathrm{i}\sigma_3mM_-$. 
	Then, Eq.~(\ref{eq:kernelKdiff2}) is written as
	\begin{align}
		(\partial_x+\partial_y)K_1(x,y)&=U(x)K_2(x,y)+K_2(x,y)M, \\
		(\partial_x-\partial_y)K_2(x,y)&=U(x)K_1(x,y)-K_1(x,y)M.
	\end{align}
	Integrating these equations and fixing the integration constants to realize the boundary conditions $ K_2(x,x)=\frac{1}{2}(U(x)-M) $ and $ \lim_{y\rightarrow-\infty}K(y+t,y)=0 \ (t\ge 0) $ yield 
	the integral equation
	\begin{align}
		K_1(x,y)&=\int_{-\infty}^x\mathrm{d}s\left[ U(s)K_2(s,s+y-x)+K_2(s,s+y-x)M \right], \\
		K_2(x,y)&=\frac{1}{2}\left( U\left( \frac{x+y}{2} \right)-M \right)+\int_{\frac{x+y}{2}}^x\mathrm{d}s\left[ U(s)K_1(s,x+y-s)-K_1(s,x+y-s)M \right].
	\end{align}
	Henceforth we show that the solution of this equation constructed by iteration converges absolutely. 
	Let $ \lVert X \rVert=(\operatorname{tr}X^\dagger X)^{1/2} $ be the Frobenius norm. 
	We define $ K_1^{(n)},\ K_2^{(n)} $ by the recurrence relation
	\begin{align}
		K_2^{(0)}(x,y)&=\frac{1}{2}\left( U\left( \frac{x+y}{2} \right)-M \right), \\
		K_1^{(n)}(x,y)&=\int_{-\infty}^x\mathrm{d}s\left[ U(s)K_2^{(n)}(s,s+y-x)+K_2^{(n)}(s,s+y-x)M \right] \quad (n\ge0), \\
		K_2^{(n+1)}(x,y)&=\int_{\frac{x+y}{2}}^x\mathrm{d}s\left[ U(s)K_1^{(n)}(s,x+y-s)-K_1^{(n)}(s,x+y-s)M \right] \quad (n\ge0).
	\end{align}
	We assume that there exist real $ a,b >0 $ such that
	\begin{align}
		\left\Vert \frac{1}{2}\left( U(x)-M \right)  \right\Vert \le a\mathrm{e}^{bx} \label{eq:kernelKapp005}
	\end{align}
	for any real $ x $. This inequality is too rough an estimate for $ x\gg 0 $, since  $ U(x)-M $ comes close to a constant matrix $ U(+\infty)-M $ in the limit $ x\rightarrow+\infty $. It is, however, sufficient to show the boundedness of the solution. We also define $ w(x):= 2\max_{-\infty\le s\le x} \Vert U(s) \Vert, $ 
	which is obviously non-decreasing and real and positive for any real $ x $, since $ w(x)\ge 2\lVert M \rVert>0  $. By induction,  we can soon prove  $ \Vert K_1^{(n)}(x,y) \Vert \le w(x)^{2n+1}\frac{1}{n!}\left( \frac{x-y}{2} \right)^n\frac{a}{b^{n+1}}\mathrm{e}^{b\frac{x+y}{2}} $ and $ \Vert K_2^{(n)}(x,y) \Vert \le w(x)^{2n}\frac{1}{n!}\left( \frac{x-y}{2} \right)^n\frac{a}{b^n}\mathrm{e}^{b\frac{x+y}{2}} $ (note: $ x-y\ge 0 $ by definition).  
	Therefore
	\begin{align}
		\Vert K_1(x,y) \Vert &\le \sum_{n=0}^\infty \Vert K_1^{(n)}(x,y) \Vert \le w(x)\frac{a}{b}\mathrm{e}^{b\frac{x+y}{2}}\mathrm{e}^{\frac{w(x)^2(x-y)}{2b}}, \label{eq:kernelKapp003} \\
		\Vert K_2(x,y) \Vert &\le \sum_{n=0}^\infty \Vert K_2^{(n)}(x,y) \Vert \le a\mathrm{e}^{b\frac{x+y}{2}}\mathrm{e}^{\frac{w(x)^2(x-y)}{2b}}, \label{eq:kernelKapp004}
	\end{align}
	which shows the absolute convergence. Note that the proof given here is a minimal one in the sense that we only show the boundedness of the solution and the inequalities (\ref{eq:kernelKapp003}) and (\ref{eq:kernelKapp004}) do not represent a correct asymptotic behavior of $ K(x,y) $ for large $ x,y $. To obtain a more refined estimate, we should modify the inequality (\ref{eq:kernelKapp005}). \\
	\indent Thus, the existence of the integral representation of the Jost function (\ref{eq:JostkernelK}) follows.

\section{Existence and uniqueness of the solution of the GLM equation}\label{app:glmuniq}
	In this appendix, we show the existence and uniqueness of the solution of the GLM equation. Let $ x\in\mathbb{R} $ be a fixed real parameter. Let $ q(y)=(q_1(y),\dots,q_{2d}(y)) $ be a  $ 2d $-component row-vector function defined in the interval $ (-\infty,x) $. We assume that it is square-integrable: $ \sum_{j=1}^{2d} \int_{-\infty}^x\mathrm{d}y |q_j(y)|^2 <\infty $. Using the kernel $ \Omega $ appearing in the GLM equation (\ref{eq:GLM00}), we define the integral operator  $ \mathcal{T} $ by
	\begin{align}
		\mathcal{T}[q](y):=\int_{-\infty}^x\mathrm{d}z q(z)\Omega(z,y).
	\end{align}
	We also assume $ \int_{-\infty}^x\mathrm{d}z\int_{-\infty}^x\mathrm{d}y \Vert \Omega(z,y) \Vert^2 <\infty $, 
	which is basically satisfied for a reflection coefficient $ R(s) $ obtained from a potential $ \Delta(x) $ rapidly coming close to a constant value for $ x\rightarrow\pm\infty $. With this assumption, $ \mathcal{T} $ becomes a Hilbert-Schmidt operator hence compact. Therefore, by the Fredholm alternative (or the Riesz-Schauder theorem), if we can prove
	\begin{align}
		(I+\mathcal{T})[q](y)=0 \quad \rightarrow \quad q(y)=0, \label{eq:appglm01}
	\end{align}
	then the existence and uniqueness of the solution of the GLM equation follows. \\
	\indent Let us prove (\ref{eq:appglm01}). In the equation $ (I+\mathcal{T})[q](y)=0 $, multiplying $ q(y)^\dagger $ from right and integrating it from $ -\infty $ to $ x $ gives
	\begin{align}
		0=\int_{-\infty}^x\mathrm{d}y\left[ q(y)q(y)^\dagger + \int_{-\infty}^x\mathrm{d}z q(z)\Omega(z,y)q(y)^\dagger \right]=I_1+I_2+I_3,
	\end{align}
	where
	\begin{align}
		I_1&=\left\Vert \int_{-\infty}^x\mathrm{d}z q(z)W(z) \right\Vert^2, \\
		I_2&=\frac{m}{8\pi}\int_{-\infty}^\infty\frac{\mathrm{d}s}{s^2}\left\Vert \int_{-\infty}^x\mathrm{d}z q(z)\left[ \Phi_-(z,s)\Delta_-^\dagger+s\Phi_-(z,s^{-1})R(s) \right] \right\Vert^2, \\
		I_3&=\frac{m}{8\pi}\int_{-\infty}^\infty\frac{\mathrm{d}s}{s^2}\left\Vert \int_{-\infty}^x\mathrm{d}zq(z) s\Phi(z,s^{-1})\Delta_-^\dagger T(s^{-1})^\dagger \right\Vert^2.
	\end{align}
	Here, the expression for $ I_2+I_3 $ can be found as follows: We can derive the relation
	\begin{align}
		\delta(z-y)+\Omega_{\text{sc}}(z,y)&=\frac{m}{8\pi}\int_{-\infty}^\infty\frac{\mathrm{d}s}{s^2}\left[ \Phi_-(z,s)\Delta_-^\dagger +s\Phi_-(z,s^{-1})R(s)  \right]\left[ \Phi_-(y,s)\Delta_-^\dagger+s\Phi_-(y,s^{-1})R(s) \right]^\dagger\nonumber \\
		&\qquad +\frac{m}{8\pi}\int_{-\infty}^\infty\frac{\mathrm{d}s}{s^2}\left[ s\Phi_-(z,s^{-1}) \right]\left( I_d-R(s)R(s)^\dagger \right)\left[ s\Phi_-(y,s^{-1}) \right]^\dagger
	\end{align}
	by combining the completeness relation (\ref{eq:complenetess00}) and the change of the integration variable $ s\rightarrow s^{-1} $. Using this and Eqs. (\ref{eq:scatinv1501}) and (\ref{eq:scatinv202}) for real $ s=s^* $, we obtain $ I_2+I_3 $. Since $ I_1,I_2, $ and $ I_3 $ are non-negative, we soon conclude $ I_1=I_2=I_3=0 $. In particular, $ I_3=0 $ implies that $ \int_{-\infty}^x\mathrm{d}zq(z) s\Phi(z,s^{-1})=0 $ holds almost everywhere, hence $ q(z)=0 $.\\
	\indent We note that the proof shown here is an analog of the proof of the same problem for the Schr\"odinger operator given in Ref. \cite{TanakaDate}.
	
\section{An inner-product formula concerning the fourth-order product of scattering eigenfunctions}\label{app:fourortho}
	In this appendix, we prove an inner-product formula for the product of four scattering eigenstates, arising from Eq.~(\ref{eq:constwjnew02}). Let $ s_1,s_2\in\mathbb{R} $, and let $ \mathcal{O} $ be an $x$-independent (but possibly dependent on $s_1,s_2$)  $ 2d\times 2d $ matrix. We want to prove: 
	\begin{align}
		&\int_{-\infty}^\infty\mathrm{d}x Y_+(x,s_1)^\dagger\left[  Y_-(x,s_2)\mathcal{O}Y_-(x,s_2)^\dagger ,\sigma_3 \right]Y_+(x,s_1)= \nonumber \\
		&\frac{\mathrm{i}s_1s_2(s_1s_2-1)}{m}\Bigg[ \frac{\operatorname{\text{p.v.}}}{s_1-s_2}\left( \sigma_3\big\{S_2^\dagger\sigma_3\mathcal{O}\sigma_3S_2,\sigma_3\big\}-S_1^\dagger\sigma_3\big\{\mathcal{O},\sigma_3\big\}S_1 \right) \nonumber \\
		&\qquad\qquad+\mathrm{i}\pi\delta(s_2-s_1)\left( [S_2^\dagger\sigma_3\mathcal{O}\sigma_3S_2,\sigma_3]-S_1^\dagger[\mathcal{O},\sigma_3]S_1 \right) \Bigg] \nonumber \\
		&+\frac{\mathrm{i}s_2(s_1-s_2)}{m}\Bigg[ \frac{\operatorname{\text{p.v.}}}{s_2-s_1^{-1}}\left( M_+\sigma_3\big\{\sigma_3,S_2^\dagger\sigma_3\mathcal{O}\sigma_3S_2\big\}M_+-S_1^\dagger M_-\sigma_3\big\{\sigma_3,\mathcal{O}\big\}M_-S_1 \right)  \nonumber \\
		&\qquad\qquad+\mathrm{i}\pi\delta(s_2-s_1^{-1})\left( M_+[\sigma_3,S_2^\dagger\sigma_3\mathcal{O}\sigma_3S_2]-S_1^\dagger M_-[\sigma_3,\mathcal{O}]M_-S_1 \right) \Bigg], \label{eq:ortho007}
	\end{align}
	where p.v. denotes Cauchy's principal value, and $ S_i=S(s_i) $. If $ \mathcal{O} $ satisfies $ \{\sigma_3,\mathcal{O}\}=\{\sigma_3,S_2^\dagger\mathcal{O}S_2\}=0 $, the p.v. part vanishes. This is the case of Subsec.~\ref{subsec:solvegap}, Eq.~(\ref{eq:ortho00701}). \\
	\indent In the following calculation, we regard $ s_1 $ as a constant and $ s_2 $ as an integration variable. We write $ k_i=k(s_i),\ \epsilon_i=\epsilon(s_i) $. The relation (\ref{eq:constwjnew02}) gives
	\begin{align}
		&\int_{-L}^L\mathrm{d}x Y_+(x,s_1)^\dagger\left[  Y_-(x,s_2)\mathcal{O}Y_-(x,s_2)^\dagger ,\sigma_3 \right]Y_+(x,s_1) \nonumber \\
		&=\frac{[Y_+(x,s_1)^\dagger\sigma_3Y_-(x,s_2)\mathcal{O}Y_-(x,s_2)^\dagger\sigma_3Y_+(x,s_1)]_{-L}^L}{-\mathrm{i}(\epsilon_1-\epsilon_2)}.
	\end{align}
	 If $ L $ is sufficiently large, the above values at $ x=\pm L $ can be approximated by the plane-wave asymptotic form (\ref{eq:asymptoticY}). 
	For brevity we write $ \tilde{\mathcal{O}}=\sigma_3S(s_2)^\dagger\sigma_3\mathcal{O}\sigma_3S(s_2)\sigma_3 $. Then, using $ S(s)^{-1}=\sigma_3 S(s)^\dagger \sigma_3 $ for real $ s $, the above integral is
	\begin{align}
		&[Y_+(x,s_1)^\dagger\sigma_3Y_-(x,s_2)\mathcal{O}Y_-(x,s_2)^\dagger\sigma_3Y_+(x,s_1)]_{-L}^L=I_1+I_2+(\text{nonsingular terms}), \\
		I_1&:=(s_1s_2-1)^2\left[ \sigma_3 \mathrm{e}^{-\mathrm{i}\theta_-\sigma_3}\tilde{\mathcal{O}}\mathrm{e}^{\mathrm{i}\theta_-\sigma_3}\sigma_3-S(s_1)^\dagger\sigma_3 \mathrm{e}^{\mathrm{i}\theta_-\sigma_3}\mathcal{O}\mathrm{e}^{-\mathrm{i}\theta_-\sigma_3}\sigma_3S(s_1) \right], \\
		I_2&:=(s_1-s_2)^2\left[ \sigma_3M_+\mathrm{e}^{\mathrm{i}\theta_+\sigma_3}\tilde{\mathcal{O}}\mathrm{e}^{-\mathrm{i}\theta_+\sigma_3}M_+\sigma_3-S(s_1)^\dagger\sigma_3M_-\mathrm{e}^{-\mathrm{i}\theta_+\sigma_3}\mathcal{O}\mathrm{e}^{\mathrm{i}\theta_+\sigma_3}M_-\sigma_3S(s_1) \right],
	\end{align}
	where $ \theta_\pm:=(k_1\pm k_2)L $. The nonsingular terms are proportional to $ (\epsilon_1-\epsilon_2)\mathrm{e}^{\pm2\mathrm{i}k_1 L} $ or $ (\epsilon_1-\epsilon_2)\mathrm{e}^{\pm2\mathrm{i}k_2 L} $, and hence, these terms are finite at $ \epsilon_1=\epsilon_2 $ and vanish for $ L\rightarrow\infty $. After rewriting the exponential factors in $ I_1,I_2 $ by $ \sin2\theta_\pm $ and $ \cos2\theta_\pm $ and using Eq. (\ref{eq:e1e2tok1k2}), we use distributional formulae $ \mathrm{e}^{\mathrm{i}kL}\rightarrow 0,\ \frac{\sin kL}{k}\rightarrow \pi\delta(k) $, and $ \frac{1-\cos kL}{k}\rightarrow\operatorname{\text{p.v.}}\frac{1}{k} $. 
	Rewriting the resultant expression by $ \delta(k_2\pm k_1)=\frac{2}{m(1+s_2^{-2})}\big[\delta(s_2-s_1^{\mp1})+\delta(s_2+s_1^{\pm1})\big] $ and $ (s_1^2s_2^2-1)\delta(s_2+s_1^{-1})=(s_1^2-s_2^2)\delta(s_2+s_1)=0 $ (due to $ x\delta(x)=0 $), we obtain Eq.~(\ref{eq:ortho007}).

\end{document}